\documentclass[a4paper,11pt]{article}
\usepackage{graphicx} 
\usepackage{cite}
\usepackage{authblk}
\usepackage{caption}
\usepackage{xcolor}

\title{Comparison of MOND and Verlinde's emergent gravity in dwarf spheroidals}

\author[1]{Youngsub Yoon}
\author[2]{Sanghyeon Han}
\author[2,3]{Ho Seong Hwang}

\affil[1]{\emph{Department of Physics and Astronomy, Sejong University,
209 Neungdong-ro Gwangjin-gu, Seoul 05006, Republic of Korea}}
\affil[2]{\emph{Department of Physics and Astronomy, Seoul National University, 1~Gwanak-ro, Gwanak-gu, Seoul 08826, Republic of Korea}}
\affil[3]{\emph{SNU Astronomy Research Center, Seoul National University, 1 Gwanak-ro, Gwanak-gu,
Seoul 08826, Republic of Korea}}

\begin{document}
 
\maketitle
\begin{abstract}
    We apply Modified Newtonian Dynamics (MOND) and Verlinde's emergent gravity separately to calculate the radial accelerations in 23 dwarf spheroidals. Then, we compare them with the observed radial accelerations. In our earlier work, we determined that, when the data set is considered in its entirety without isolating individual dwarf spheroidal, Verlinde's emergent gravity is in close agreement with the observed values. In the present work, we additionally confirm that, for 21 of the 23 samples examined, Verlinde's emergent gravity follows the trend of the observed values within each dwarf spheroidal more closely than MOND. Combining the statistical significance of all the 23 samples, ranging from $-0.25\sigma$ to 3.41$\sigma$, we conclude that Verlinde's emergent gravity is favored over MOND at 5.2$\sigma$.
\end{abstract}
\section{Introduction}
The missing mass problem (for a review, see \cite{Trippe}), first noted about a century ago and widely recognized roughly half a century later, remains unresolved to this day. The mainstream scientific approach to this problem posits the existence of dark matter—an invisible, non-baryonic substance that does not interact with photons. According to this view, dark matter provides additional gravitational attraction that was not accounted for when the total gravitational field was calculated using the Newton-Einstein theory of gravity. Many candidates for dark matter have been proposed (for review, see \cite{darkmatter1, darkmatter2}), and many of them have been excluded by experiments. So far, the effort to directly detect dark matter has been fruitless \cite{XENON:2023cxc,LZ:2022lsv, COSINE-100:2022dvc, COSINE-100:2025kbw}.

In 1983, Milgrom proposed a novel approach to address the missing mass problem \cite{MOND1, MOND2, MOND3}. Instead of postulating dark matter, he suggested that it was the Newton–Einstein theory of gravity itself that needed to be modified. Observing that the rotation curves of galaxies are always asymptotically flat—despite large variations in their characteristics, such as luminosity, size, and surface brightness—he argued that it was implausible for dark matter in all galaxies to be distributed precisely in the way required to produce such curves. Thus, by modifying the Newton–Einstein theory, he was able to explain the universal flatness of galactic rotation curves without introducing any unseen matter. He named his proposal MOND (Modified Newtonian Dynamics) and made several bold predictions—such as the baryonic Tully–Fisher relation and specific properties of the rotation curves of low surface brightness galaxies, among others—all of which were later confirmed observationally.

More concretely, according to Milgrom's MOND proposal, the observed gravitational acceleration, $g$, is a function of the Newtonian gravitational acceleration, $g_{\rm bar}$, which accounts only for the baryonic mass. It is called an ``interpolating'' function and must satisfy the following asymptotic conditions:
\begin{equation}
g=g_{\rm bar}\quad (g_{\rm bar}\gg a_M), \quad g=\sqrt{a_M g_{\rm bar}}\quad (g_{\rm bar}\ll a_M),\label{twoconditions}
\end{equation}
where $a_M=1.2\times 10^{-10}\mathrm{m/s^2}$— known as Milgrom’s constant — sets the acceleration scale at which the MOND effect becomes significant. The first condition ensures that MOND reduces to Newtonian gravity when the gravitational acceleration is not tiny, while the second condition accounts for the flatness of galactic rotation curves and the baryonic Tully–Fisher relation.

In principle, the interpolating function can be determined by matching the observed acceleration to $g_{\rm bar}$, the Newtonian gravitational acceleration due to baryons. However, because of relatively large observational uncertainties, it is impossible to uniquely determine the correct form of the interpolating function among the infinitely many that satisfy Eq.~(\ref{twoconditions}). Nevertheless, a comparison of the observed radial accelerations at $\sim$ 2700 points in 153 late-type and 25 early-type galaxies with their corresponding Newtonian accelerations showed that the following interpolating function for the MOND-predicted acceleration provides an excellent fit within the large observational errors \cite{Lelli:2016cui}:
\begin{equation}
    g_{\rm MOND} = \frac{g_{\rm bar}}{1 - e^{-\sqrt{g_{\rm bar} / a_M}}} \, ,\label{RAR}
\end{equation}
where they obtained $a_M = (1.20 \pm 0.24) \times 10^{-10}\,\mathrm{m/s^2}$. The relatively large uncertainty reflects the $\sim20\%$ uncertainty in the mass-to-light ratio. The difference between the observed and the MOND-predicted radial accelerations was found to be fully consistent with observational errors and showed no correlation with any key galactic property, such as baryonic mass, radius, stellar surface density, or gas fraction (see Figs.~4 and 5 of \cite{Lelli:2016cui}). This is remarkable considering that some of these values span more than four or five orders of magnitude.

More recently, the dark matter paradigm has also been challenged by observed gravitational anomalies in wide binaries \cite{Hernandez0, Hernandez1, Hernandez2, Chae1, Chae2, Chae3, Chae4, YoonTianChae, Chae5}. A wide binary consists of two stars orbiting each other at very large separations. The observed relative accelerations of the binary components are significantly larger than those predicted by Newtonian gravity, once they fall below the MOND scale characterized by Milgrom's constant. Unlike the case of galactic rotation curves, this discrepancy cannot be attributed to the presence of dark matter, since the mass of dark matter near wide binaries is several orders of magnitude too small to noticeably affect their internal gravitational accelerations, let alone increase them. This phenomenon therefore provides a falsification of the dark matter hypothesis, challenging its account of galaxy rotation curves and lending support instead to the MOND interpretation. Also, recent examinations of the tidal tails of open star clusters show that Newtonian gravitation is ruled out, with MOND being consistent with the data \cite{Pavel1, Pavel2}.

Despite the great success of MOND, the same study  \cite{Lelli:2016cui} that found agreement between the observed and the MOND accelerations of late-type and early-type galaxies found that the observed gravitational accelerations for dwarf spheroidals are significantly larger than those predicted by Eq.~(\ref{RAR}). Another study \cite{Julio:2025vzr} reached the same conclusion, noting that it cannot be explained by tidal forces from the Milky Way—since none of the galaxies in the sample experiences strong tides—nor by the external field effect, which \emph{lowers} the MOND-predicted acceleration. 

Moreover, it remains a great weakness of MOND that the interpolating function cannot be obtained from first principles, but must be obtained only by comparing it with observations. In other words, the theory was constructed to fit the observations rather than being derived from physical and conceptual principles. This is in sharp contrast with theories such as general relativity.

Fortunately, Verlinde’s emergent gravity \cite{Verlinde, Yoon, YoongBgD}—a theory of gravity grounded in physical and conceptual principles—can also explain the flatness of galactic rotation curves and the baryonic Tully–Fisher relation without dark matter. It has been successfully tested against the observed radial accelerations in both late-type galaxies \cite{Yoongalaxy} and dwarf spheroidal systems \cite{HanHwangYoon}. The success of Verlinde’s emergent gravity in dwarf spheroidals, where MOND fails, can be attributed to the fact that the gravitational acceleration is not expressed as a simple function of the Newtonian acceleration. For the same Newtonian acceleration $g_{\rm bar}$, the accelerations predicted by Verlinde gravity for dwarf spheroidals are larger than those for late-type galaxies.

In this study, we expand on our earlier test of dwarf spheroidals in \cite{HanHwangYoon}. In our previous study, we focused on the fact that the Verlinde gravity acceleration $g_{\rm Ver}$ is closer to the observed acceleration $g_{\rm obs}$ than the MOND acceleration $g_{\rm MOND}$ is. This time, we will closely examine how $g_{\rm obs}$, $g_{\rm Ver}$, and $g_{\rm MOND}$ change within each dwarf spheroidal, and conclude that, for 21 of the 23 samples, $g_{\rm Ver}$ follows the observed acceleration $g_{\rm obs}$ more closely than $g_{\rm MOND}$ does.

The organization of this paper is as follows. In Section \ref{Verlindesemergentgravity}, we will briefly review Verlinde's emergent gravity. In Section \ref{method}, we will introduce the method we use to determine which of  $g_{\rm Ver}$ and $g_{\rm MOND}$ follows more closely the observed acceleration. In Section \ref{results}, we will present our results. In Section \ref{conclusions}, we conclude our paper.

\section{Verlinde's emergent gravity}\label{Verlindesemergentgravity}
Before proposing emergent gravity in 2017, Verlinde first suggested ``entropic gravity'' in 2011 \cite{Verlindeentropic}. According to this suggestion, gravity is merely the result of the tendency for entropy to always increase. By assuming that the concerned entropy is proportional to the enclosed area, he successfully derived Newton's gravity and Einstein's general relativity. Then, in his emergent gravity proposal six years later, he assumed that there is a volume contribution to entropy, in addition to the usual area law. This additional volume entropy is due to the horizon of the universe, which he assumed to be de Sitter. Thus, gravity deviates from the inverse-square law on the long-distance scale, where the volume contribution is no longer negligible. More precisely, it deviates in the weak acceleration regime, such as in galaxies or wide binaries.

According to Verlinde's emergent gravity, the gravitational acceleration is given by the following formula \cite{YoongBgD}: 
\begin{equation}
g_{\rm Ver}=\sqrt{g_{\rm bar}^2+g_d^2},\label{gVer}    
\end{equation}
where $g_d$, the gravitational acceleration due to ``apparent dark matter'' takes the following form \cite{Yoon}:
\begin{equation}
g_d^2=\frac{a_0}{6}(2g_{\rm bar}-\partial_r \Phi_B),\quad \Phi_B=-\frac{2 g_{\rm bar}^2}{4\pi G \rho_{\rm bar}-\partial_r g_{\rm bar}}\label{gD2}\end{equation}
where $\rho_{\rm bar}$ is the mass density of baryon. Here, $a_0=cH_0=6.7\times10^{-10}\mathrm{m/s^2}$, if our universe is de Sitter. Comparing the above equations with Eq.~(\ref{twoconditions}), we recognize that $a_0/6$ plays the role of $a_M$ in MOND. Indeed, the two values are very close, showing that Verlinde succeeded in connecting the accelerated expansion of universe with the missing mass problem, which are two separate phenomena from the perspective of MOND. 

However, our universe is not certainly de Sitter, as there are other components than the cosmological constant. Therefore, the real value of $a_0$ must be smaller. Actually, a smaller one called ``quasi-de Sitter'' value was considered in \cite{quasideSitter}. Indeed, fitting galaxy rotation curves with Verlinde's emergent gravity generalized to the absence of spherical symmetry in \cite{Yoon} shows that the quasi-de Sitter value $a_0=5.41\times 10^{-10} \mathrm{m/s^2}$ is preferred to the de Sitter value \cite{Yoongalaxy,YoonHwangcomment}. Then, $a_0/6$ no longer coincides with $a_M$. However, as the key equations of Verlinde gravity and the ones of MOND are slightly different, this is not necessarily a contradiction. Our previous investigation \cite{HanHwangYoon} shows that Verlinde's emergent gravity predicts higher accelerations in dwarf spheroidals than the MOND predictions, although $a_0/6$ for the quasi-de Sitter value is significantly smaller than $a_M$. This makes the Verlinde predictions closer to the observed ones than the MOND predictions. In the present study, we only consider the quasi-de Sitter value.

\section{Method}\label{method}

If we plot the theoretical gravitational acceleration $g_{\rm th}$ on the $x$-axis and the observed gravitational acceleration $g_{\rm obs}$ on the $y$-axis, the data will fall on the graph $y=x$, provided that the theoretical values perfectly match the observed ones. In this case, a linear regression would yield a slope of 1, corresponding to an angle of 45$^\circ$. 

In this study, we consider $g_{\rm Ver}$ and $g_{\rm MOND}$ for the theoretical gravitational accelerations by approximating the dwarf spheroidals as spheres. Thus, we have 
\begin{equation}
    g_{\rm bar}(r)=\frac{G M(r)}{r^2},
\end{equation}
where $M(r)$ is the mass of the baryon inside the sphere of radius $r$. Then, we can use Eqs.~(\ref{gVer}) and (\ref{gD2}) for $g_{\rm Ver}$, and Eq.~(\ref{RAR}) for $g_{\rm MOND}$. We will not discuss how we obtained $M(r)$ and $g_{\rm obs}$ as it is explained in detail in our previous study \cite{HanHwangYoon}. After obtaining both theoretical values, we will determine the slope of the error-weighted linearly fitted line for both theories across 23 dwarf spheroidals. The theory with a slope angle closer to 45$^\circ$ must be considered superior.

\section{Results}\label{results}

\begin{figure}
\centering

\hspace{0.4cm}\includegraphics[width=0.42\textwidth]{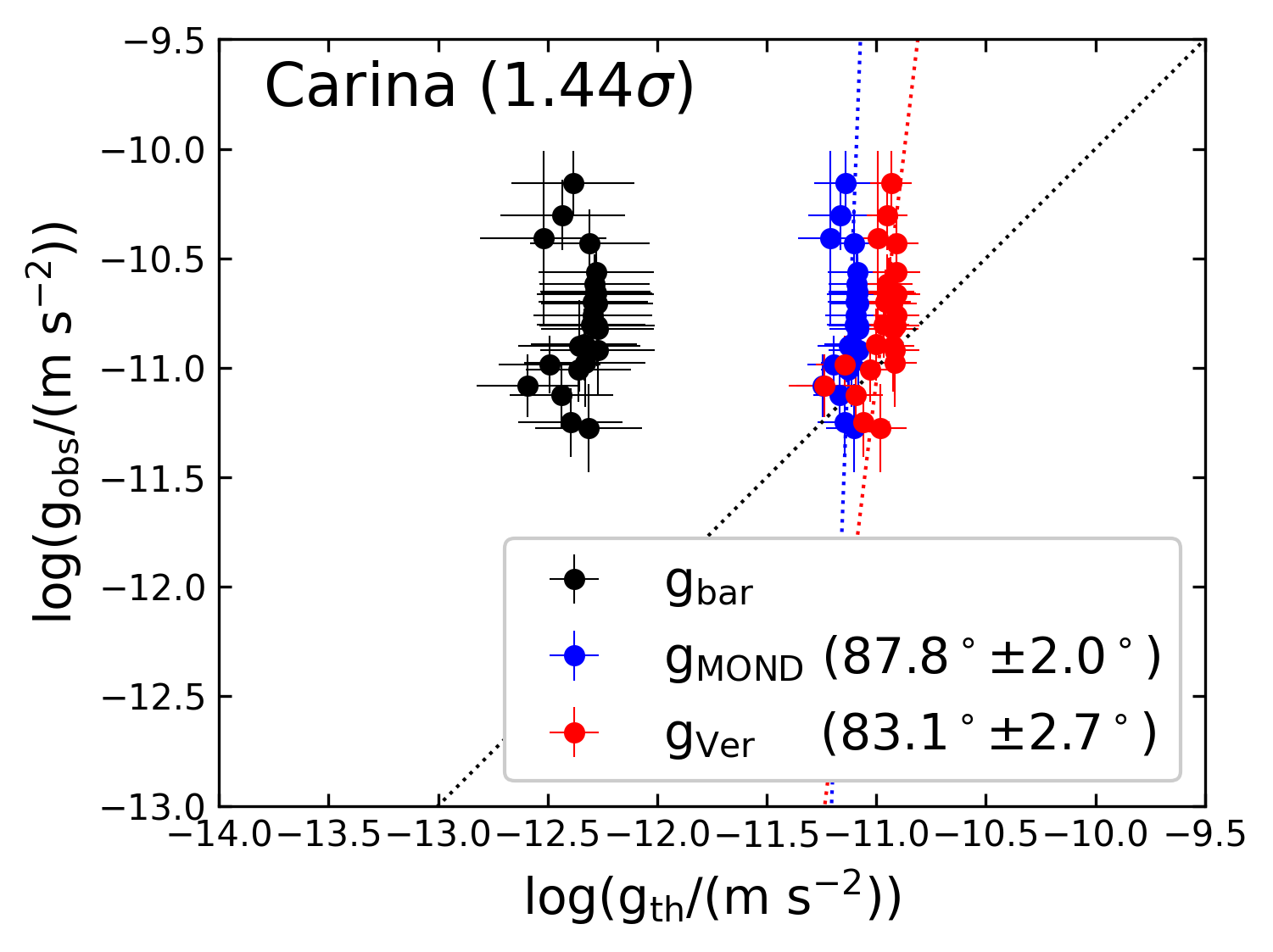} 
\includegraphics[width=0.42\textwidth]{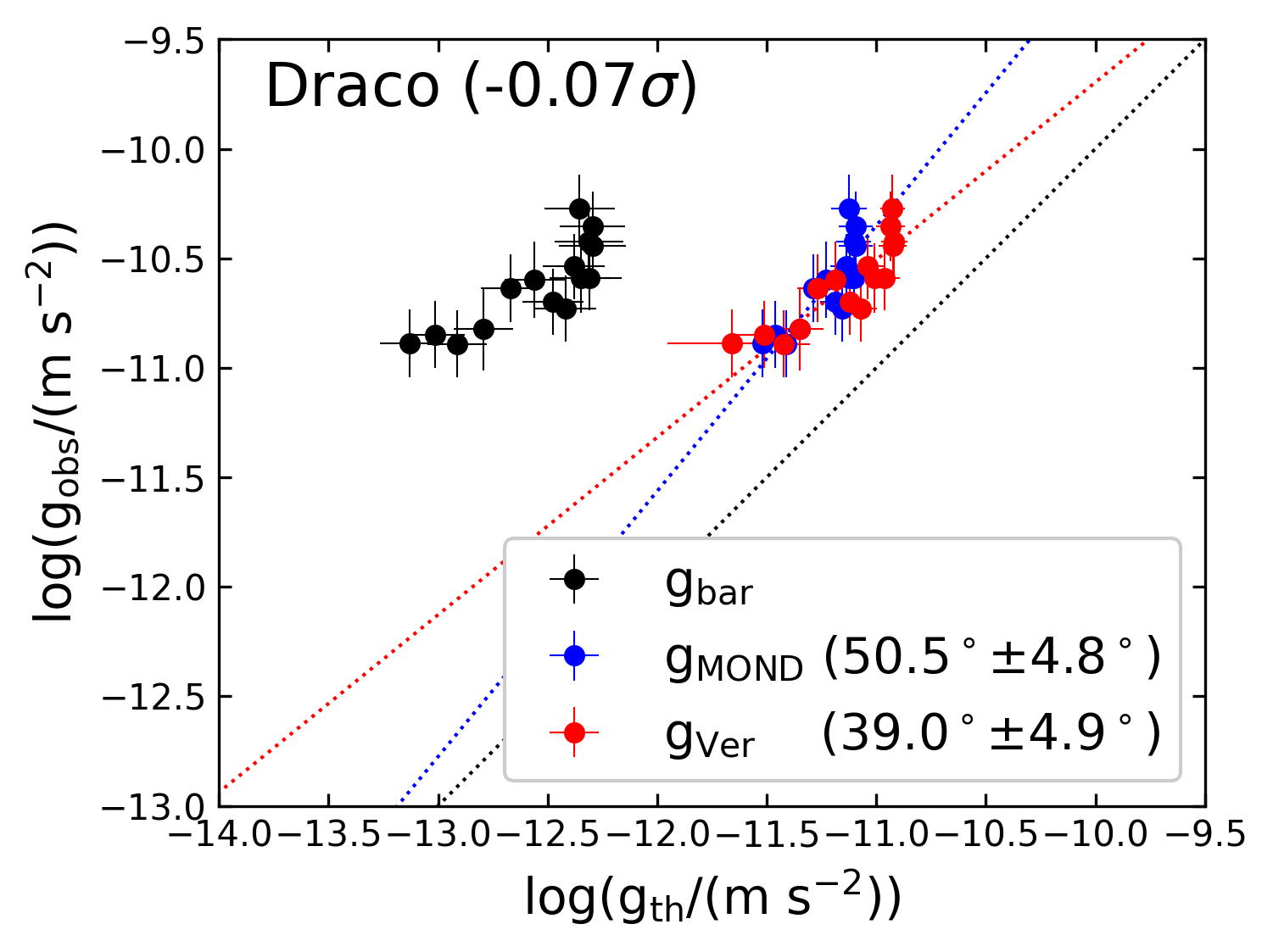} \\[1ex]

\hspace{0.4cm}\includegraphics[width=0.42\textwidth]{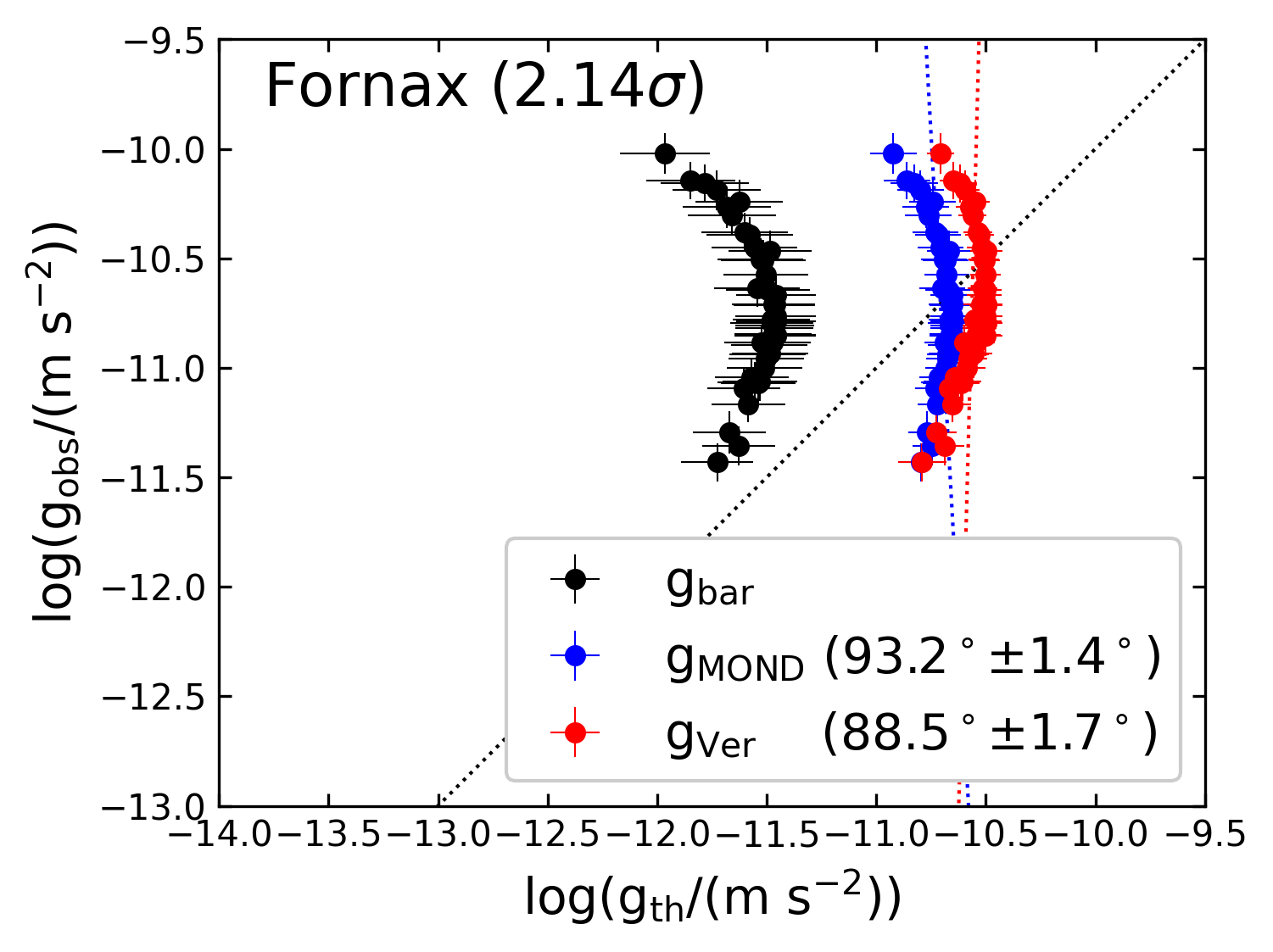} 
\includegraphics[width=0.42\textwidth]{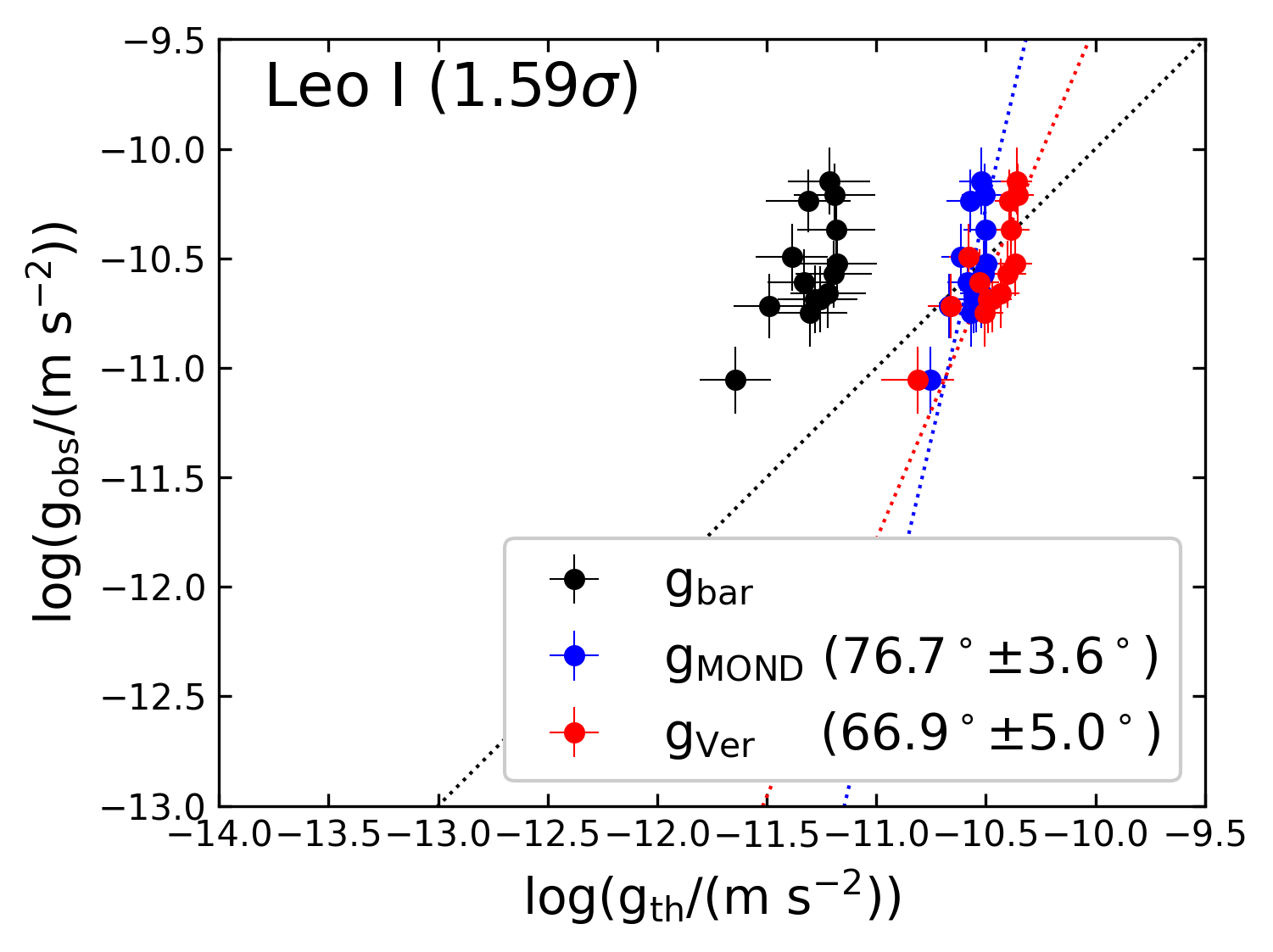} \\[1ex]

\hspace{0.4cm}\includegraphics[width=0.42\textwidth]{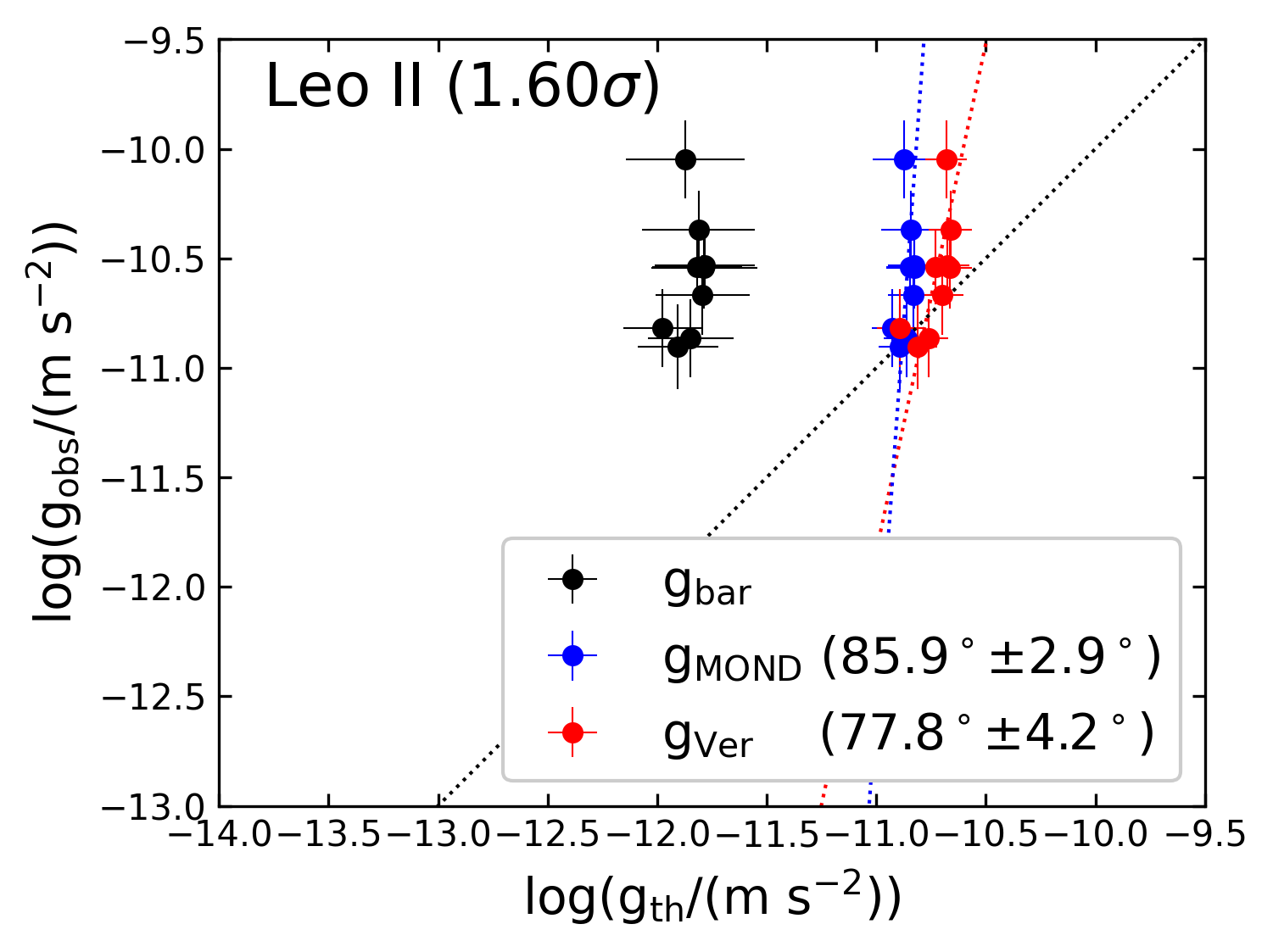} 
\includegraphics[width=0.42\textwidth]{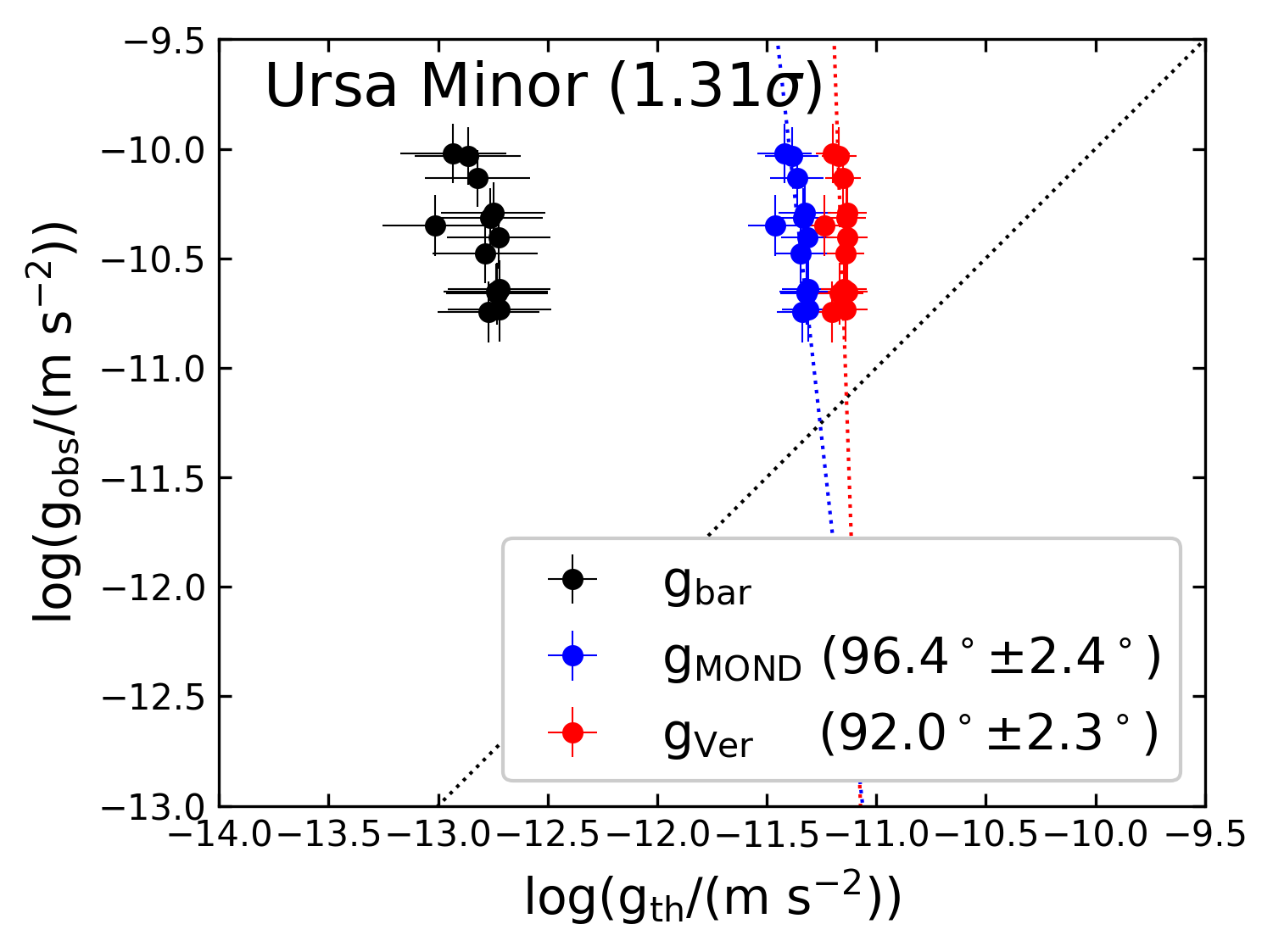} \\[1ex]

\hspace{0.4cm}\includegraphics[width=0.42\textwidth]{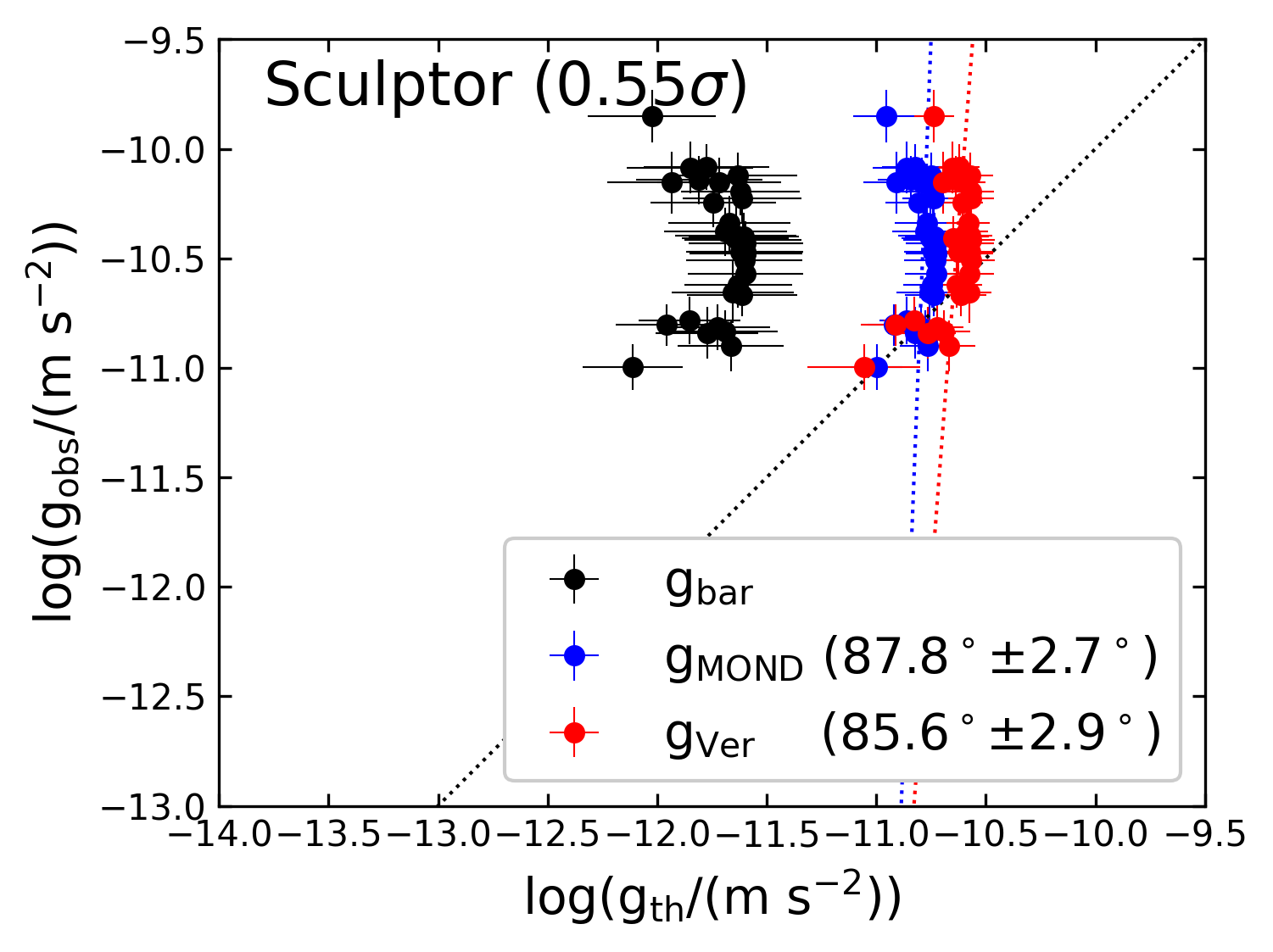} 
\includegraphics[width=0.42\textwidth]{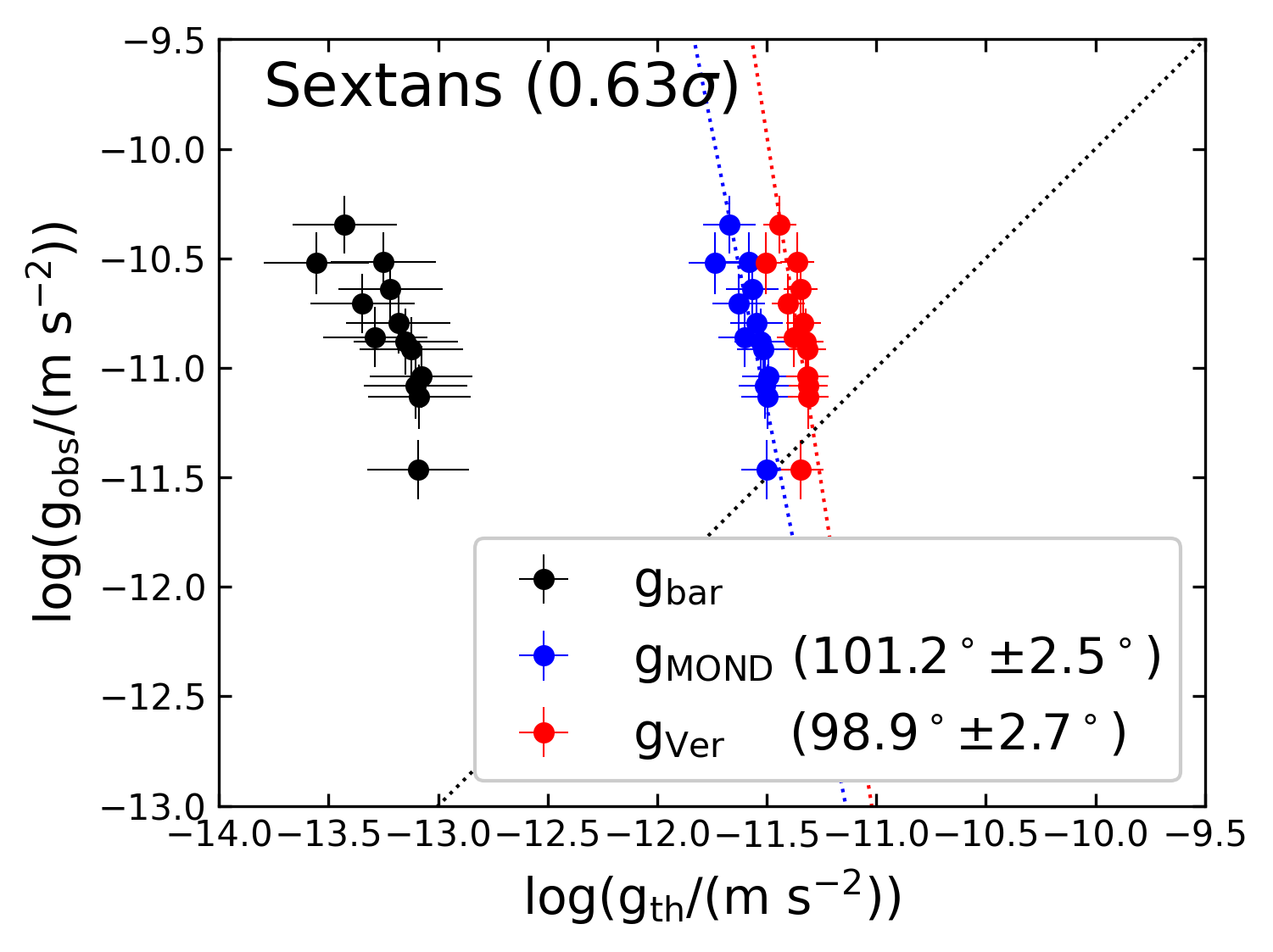}

\caption{Comparison of the theoretical gravitational accelerations with the observed accelerations for eight classical satellites of the Milky Way. The colored dotted line represents the error-weighted best fit for each data set.
For seven of the eight samples, the red dots (Verlinde gravity) fit better than the blue dots (MOND), as the slopes of the red lines (Verlinde gravity) are closer to 45$^\circ$ line than those of the blue lines (MOND).} \label{eightfigures}
\end{figure}

\begin{figure}
\centering

\hspace{0.4cm}\includegraphics[width=0.42\textwidth]{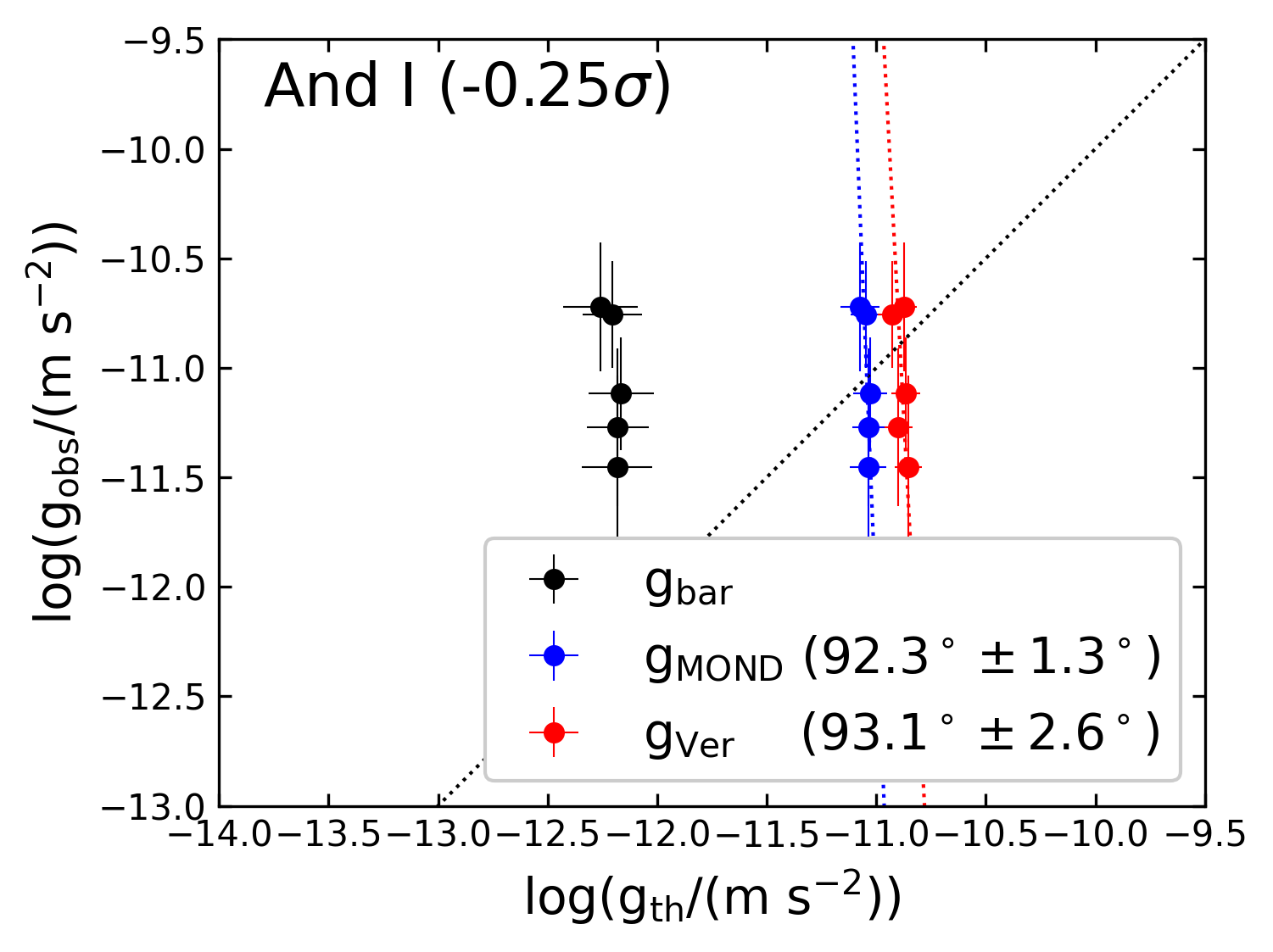} 
\includegraphics[width=0.42\textwidth]{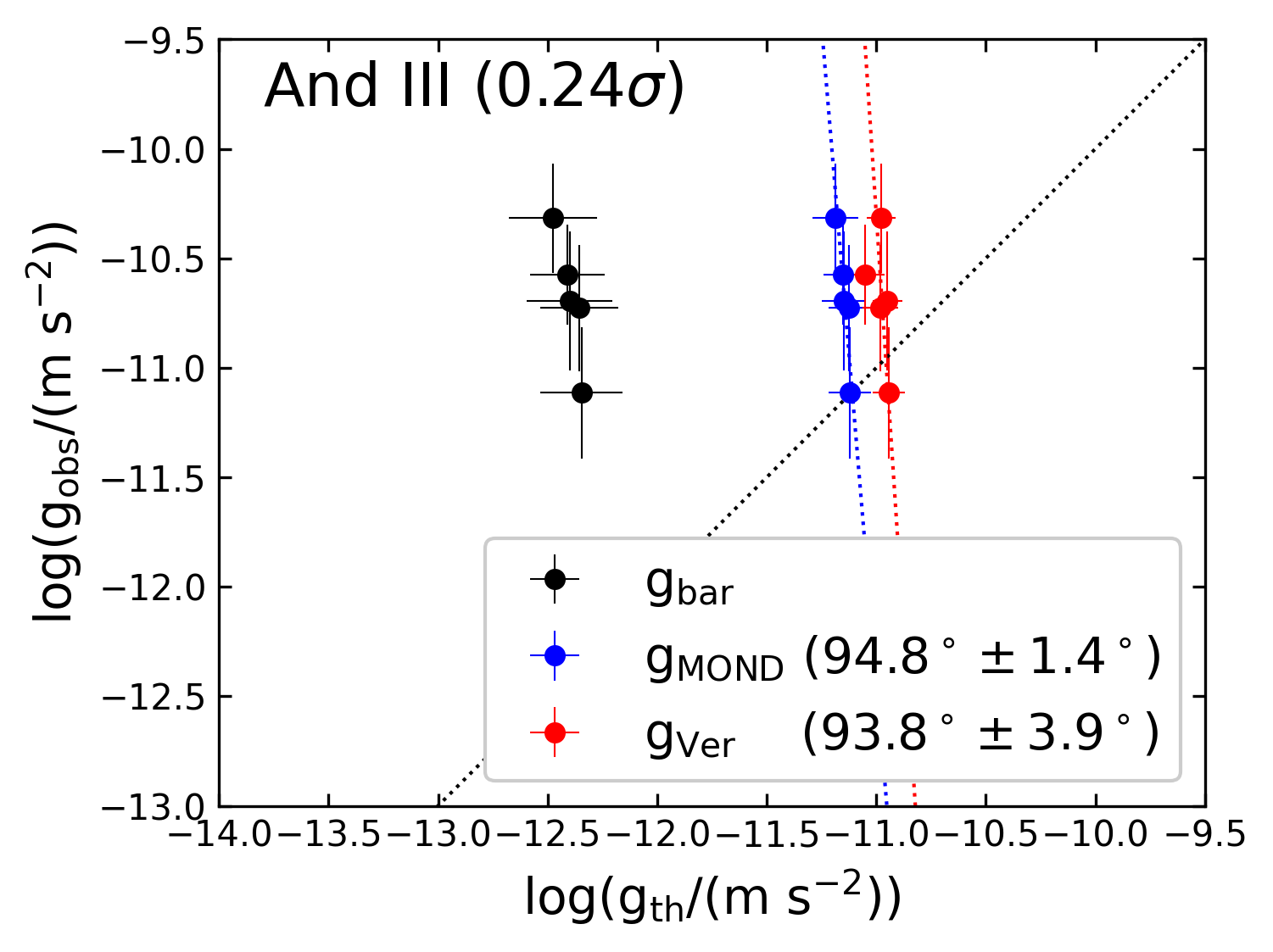} \\[1ex]

\hspace{0.4cm}\includegraphics[width=0.42\textwidth]{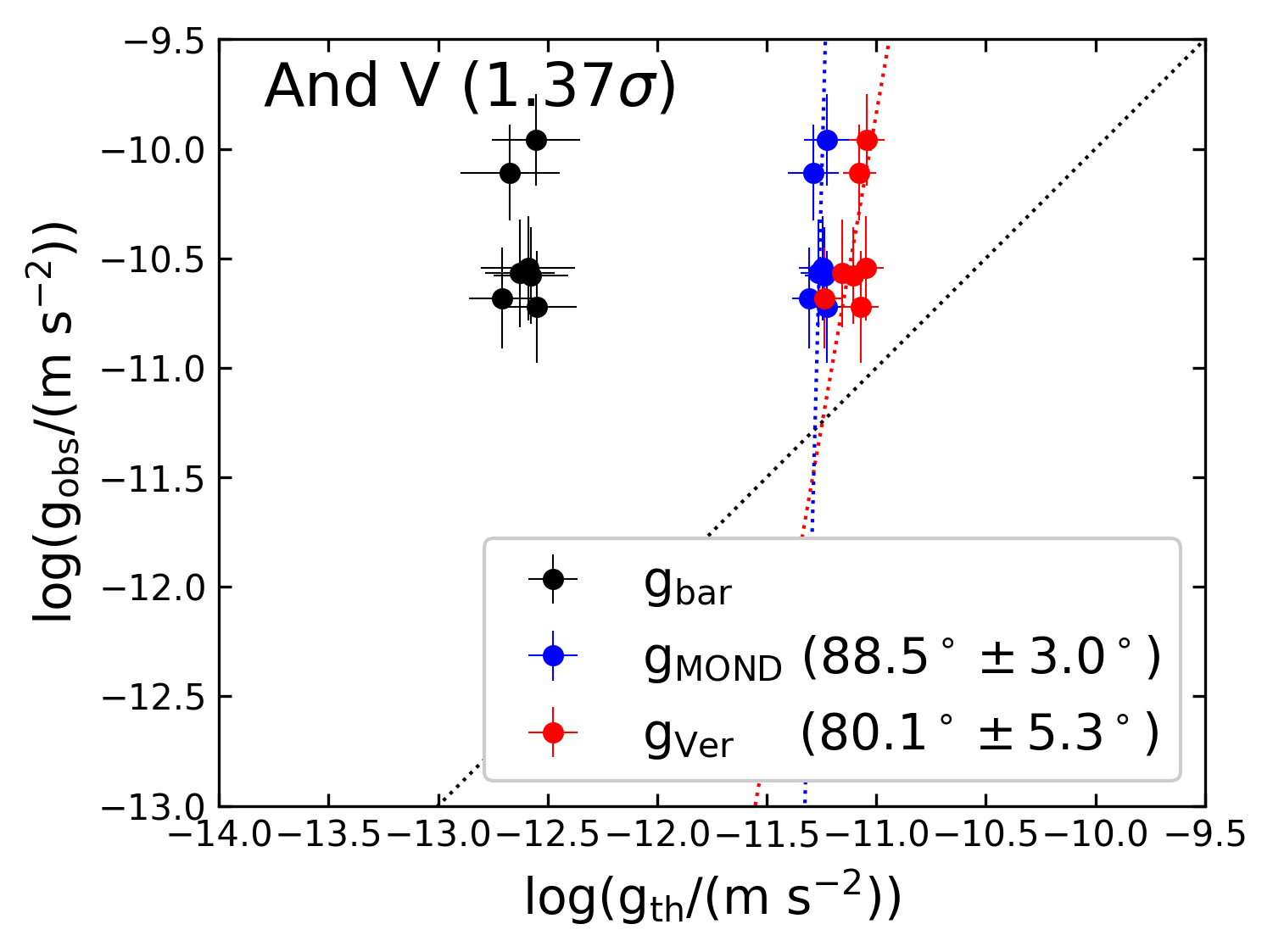} 
\includegraphics[width=0.42\textwidth]{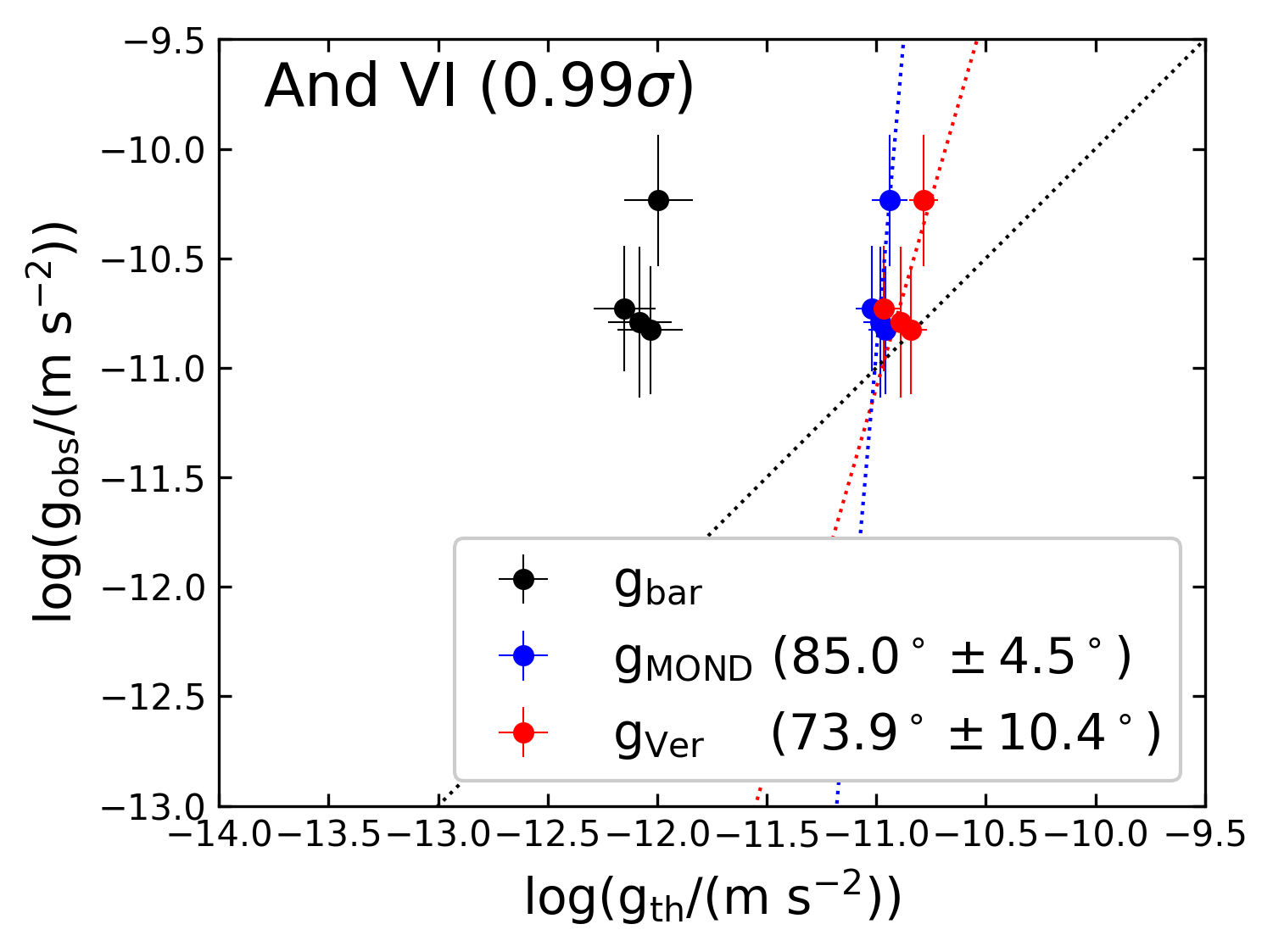} \\[1ex]

\hspace{0.4cm}\includegraphics[width=0.42\textwidth]{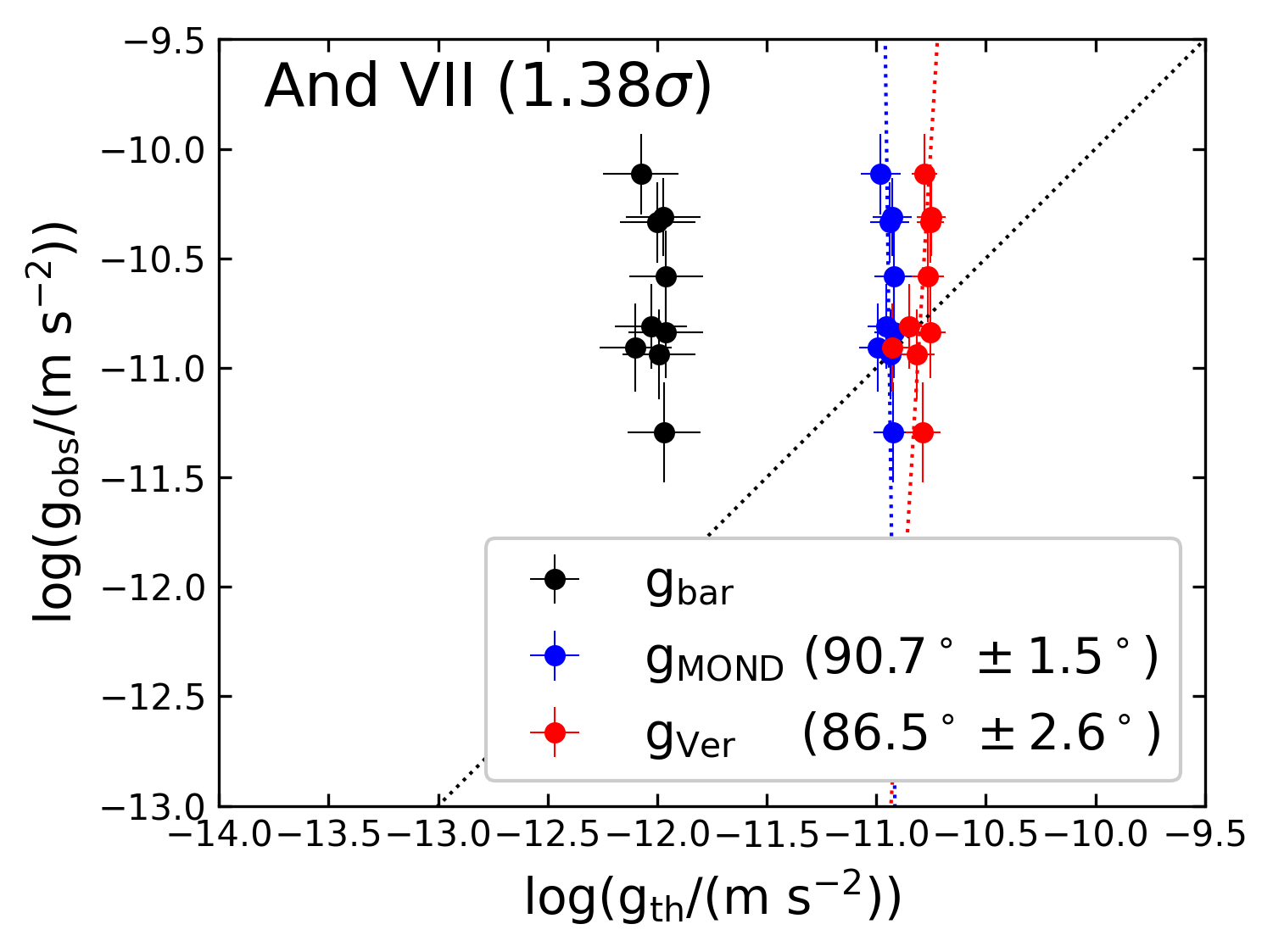} 
\includegraphics[width=0.42\textwidth]{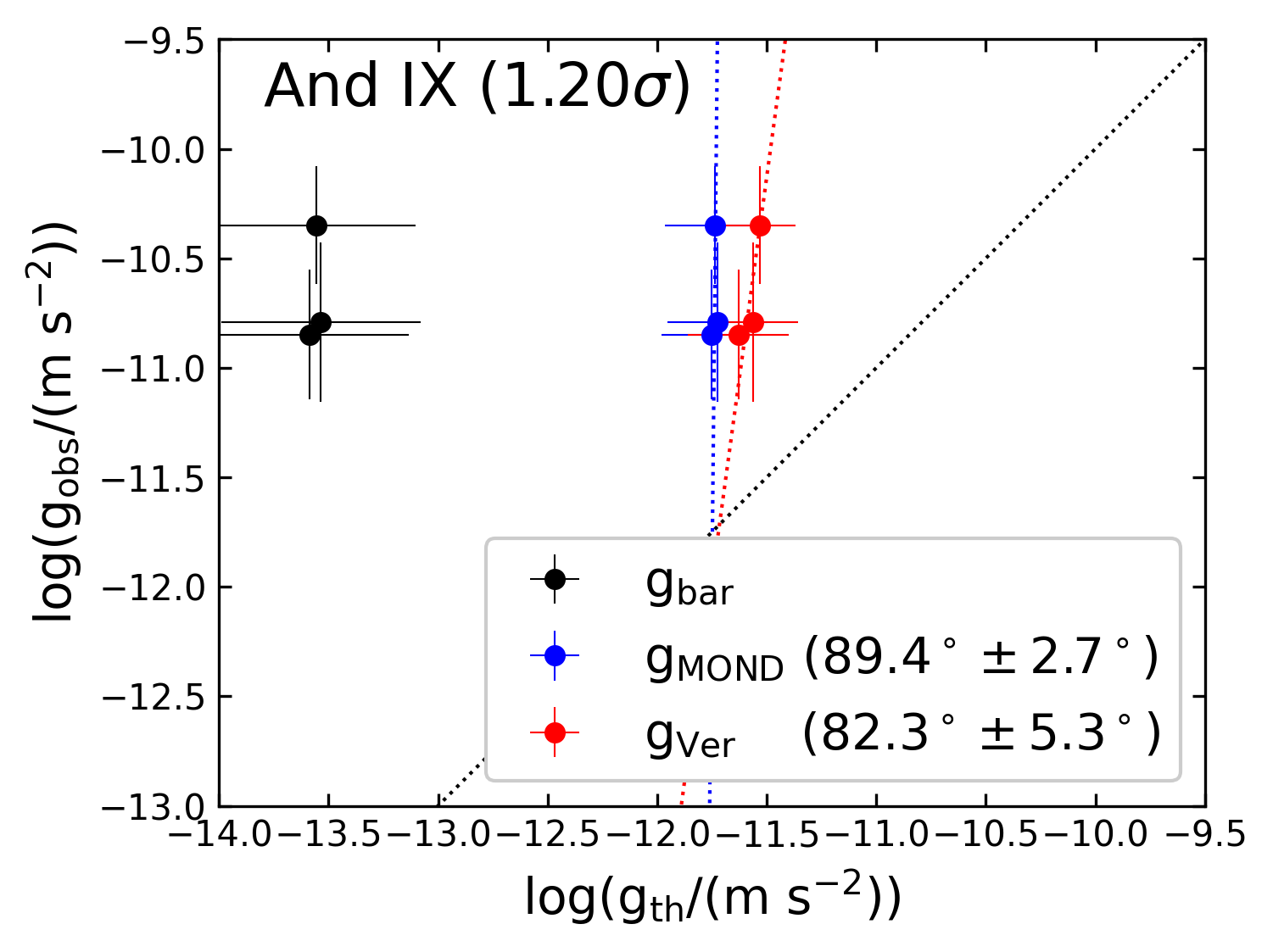} \\[1ex]

\hspace{0.4cm}\includegraphics[width=0.42\textwidth]{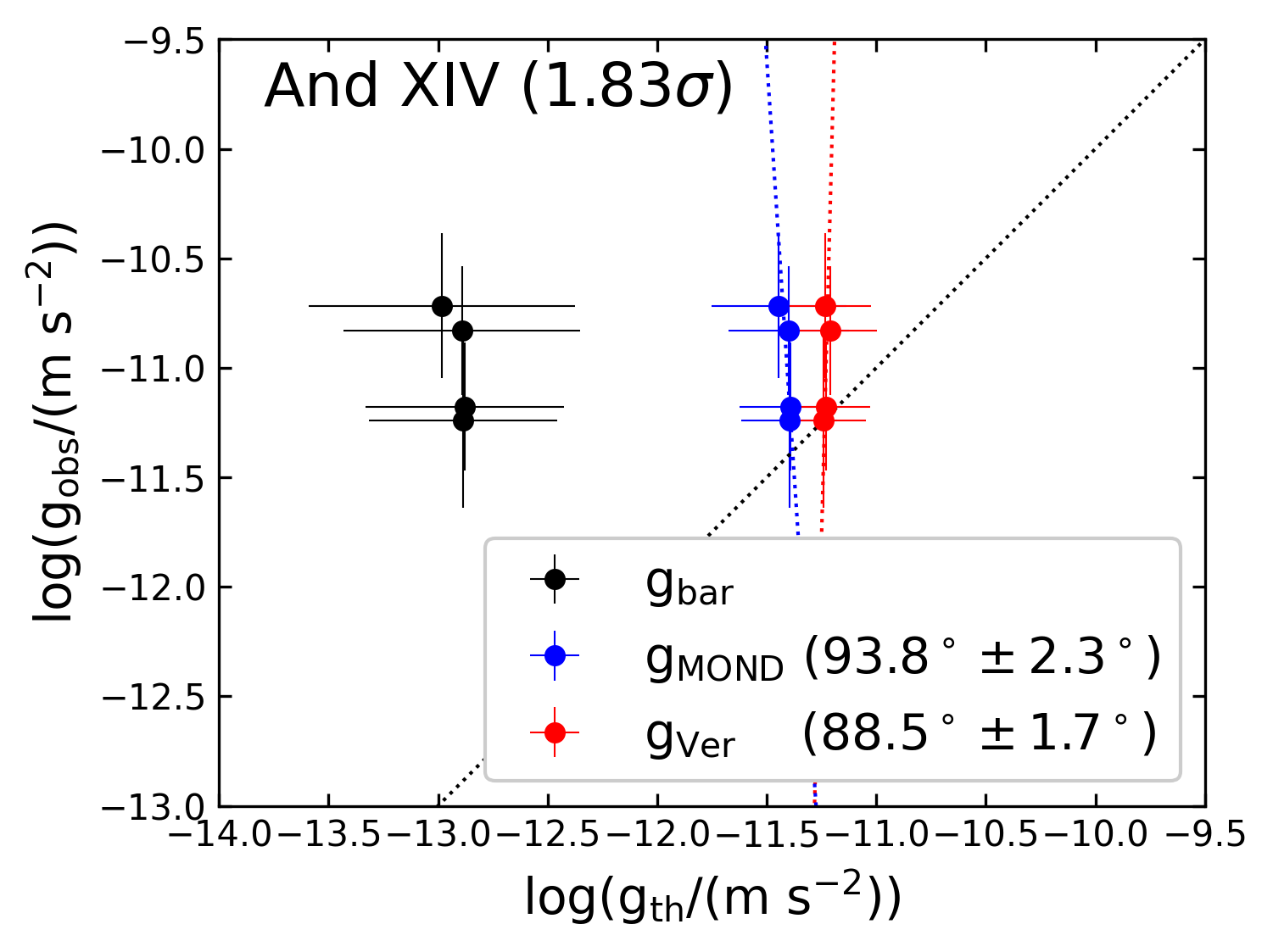} 
\includegraphics[width=0.42\textwidth]{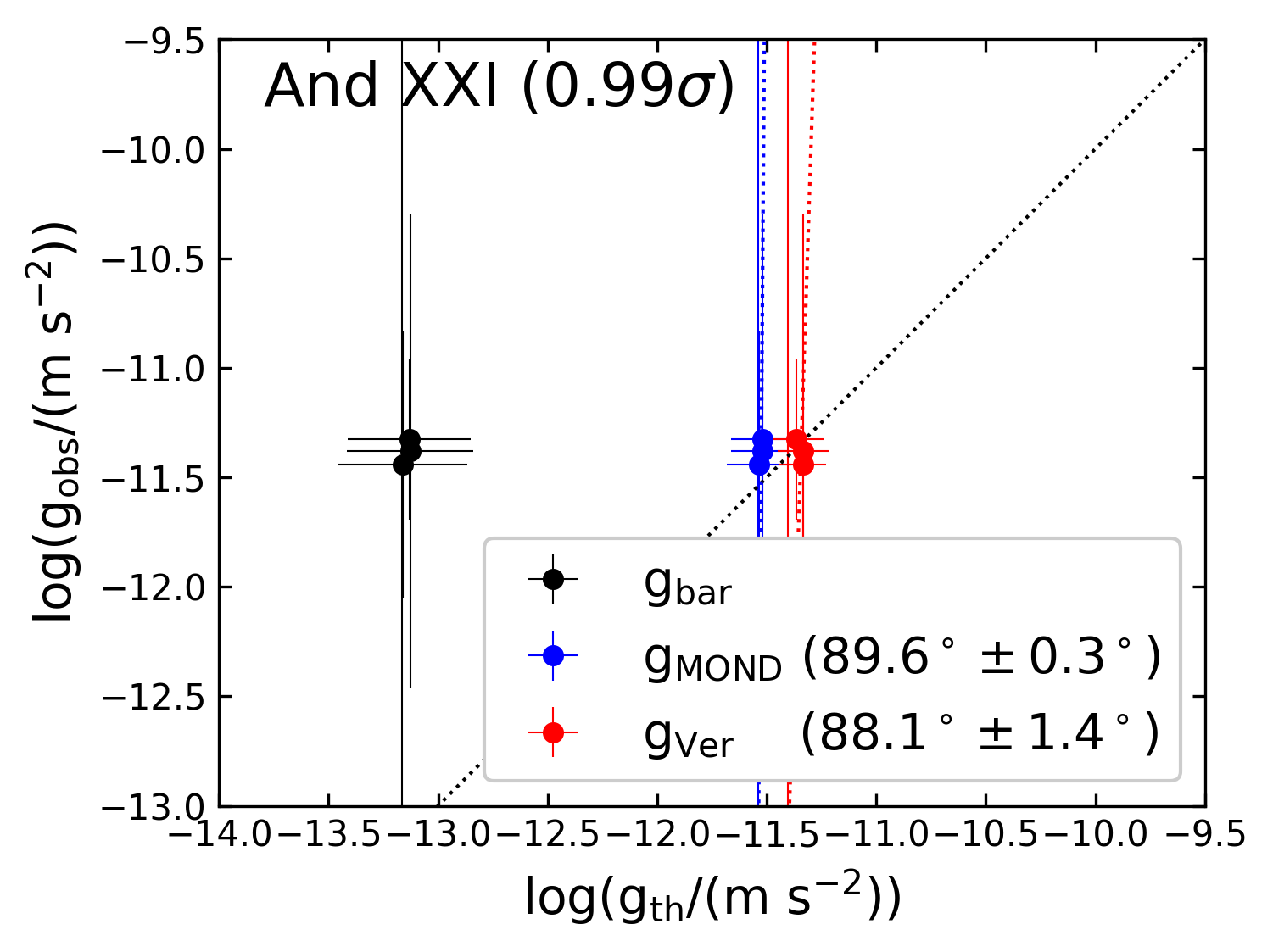}

\caption{Same as Fig. \ref{eightfigures}, but for 8 other galaxies} \label{othereightfigures}
\end{figure}

\begin{figure}
\centering

\hspace{0.4cm}\includegraphics[width=0.42\textwidth]{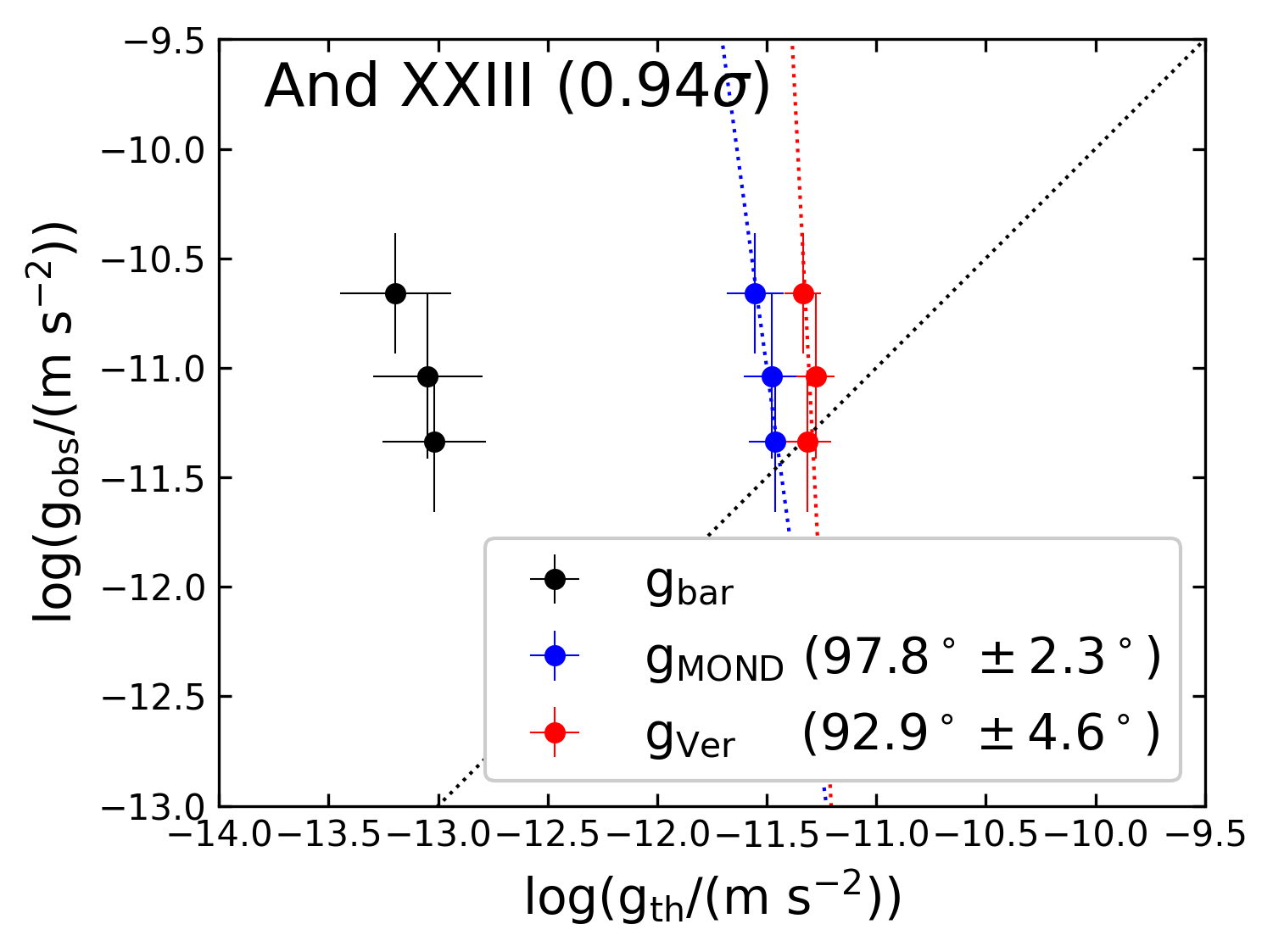} 
\includegraphics[width=0.42\textwidth]{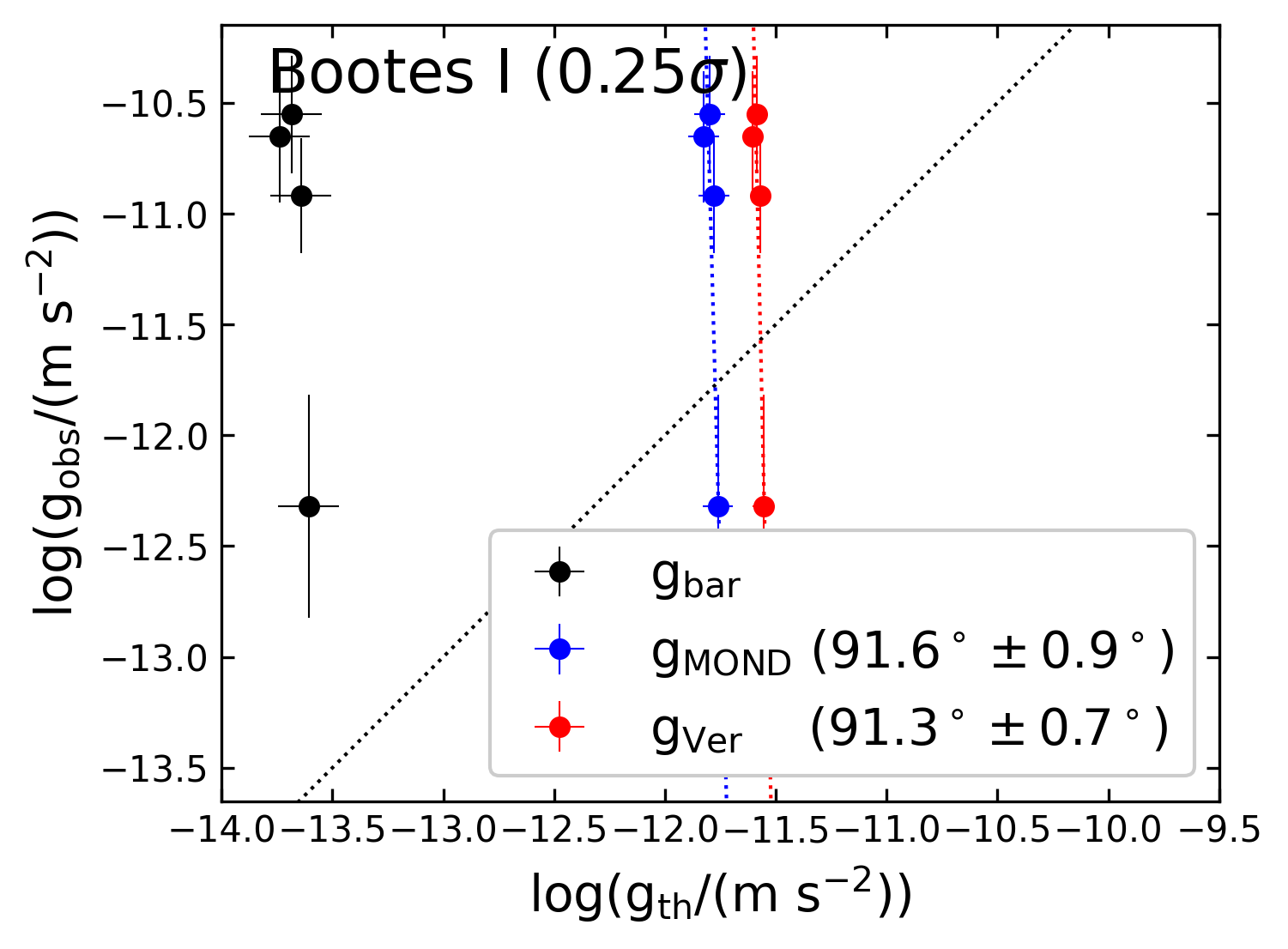} \\[1ex]

\hspace{0.4cm}\includegraphics[width=0.42\textwidth]{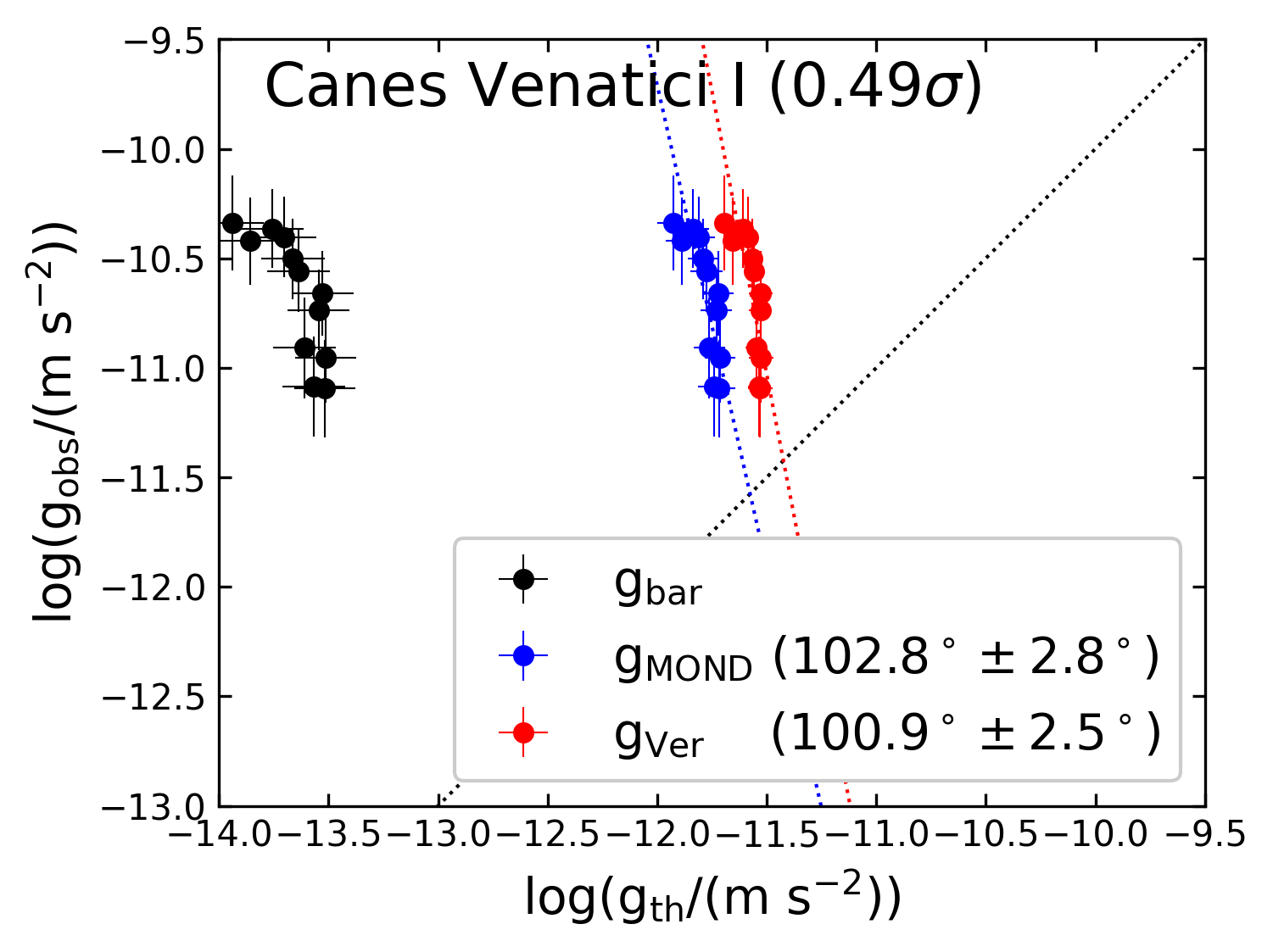} 
\includegraphics[width=0.42\textwidth]{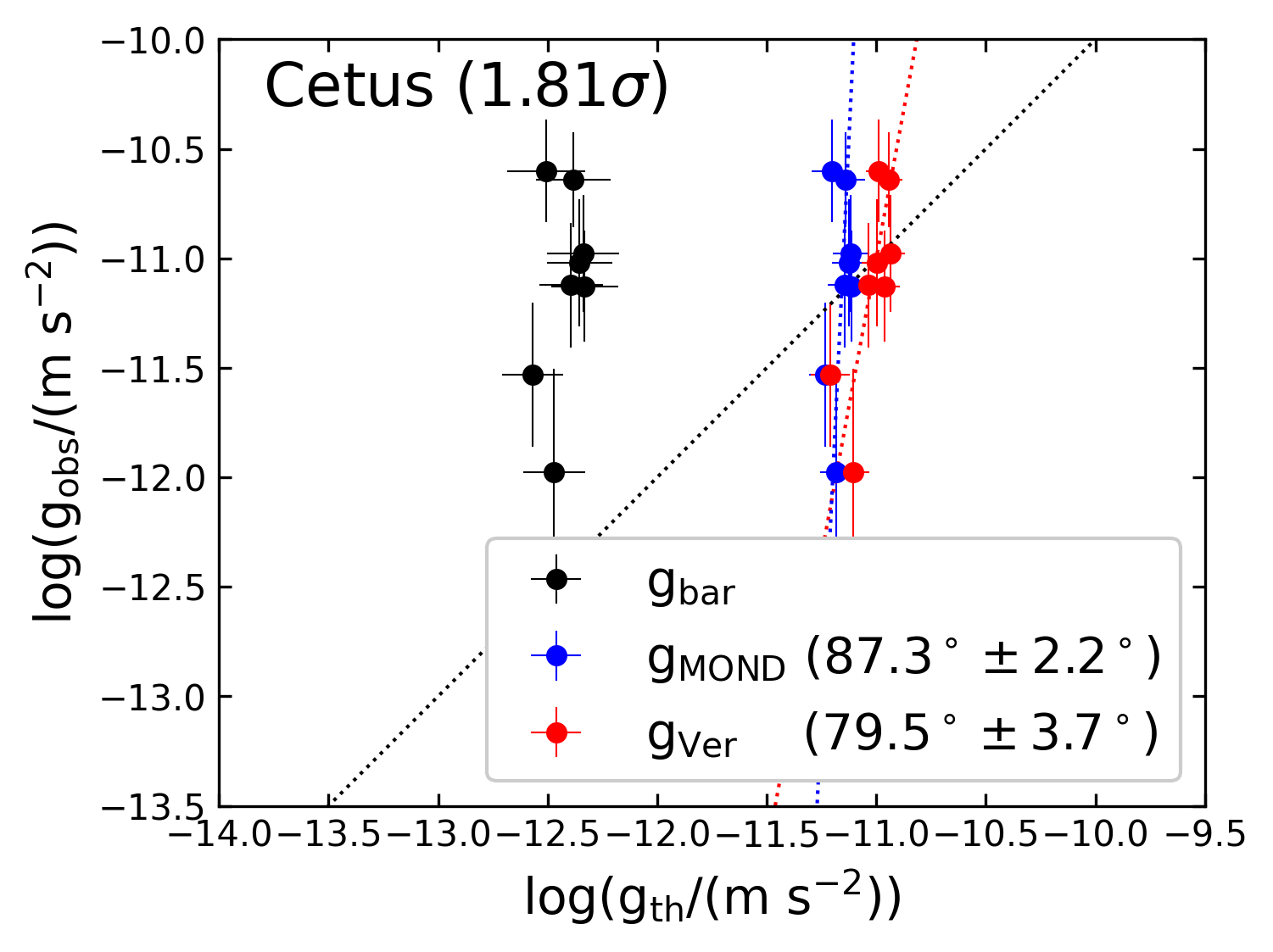} \\[1ex]

\hspace{0.4cm}\includegraphics[width=0.42\textwidth]{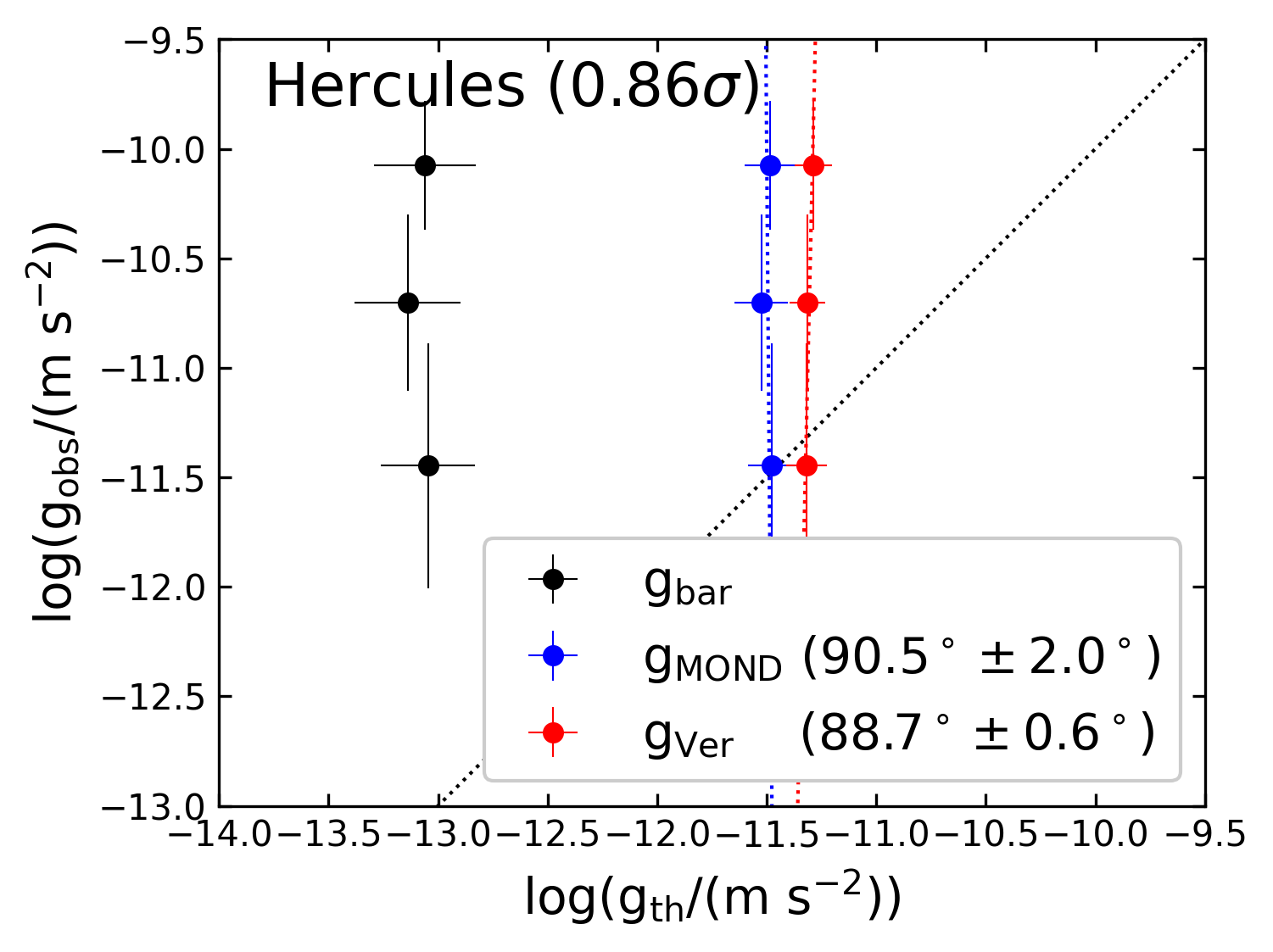} 
\includegraphics[width=0.42\textwidth]{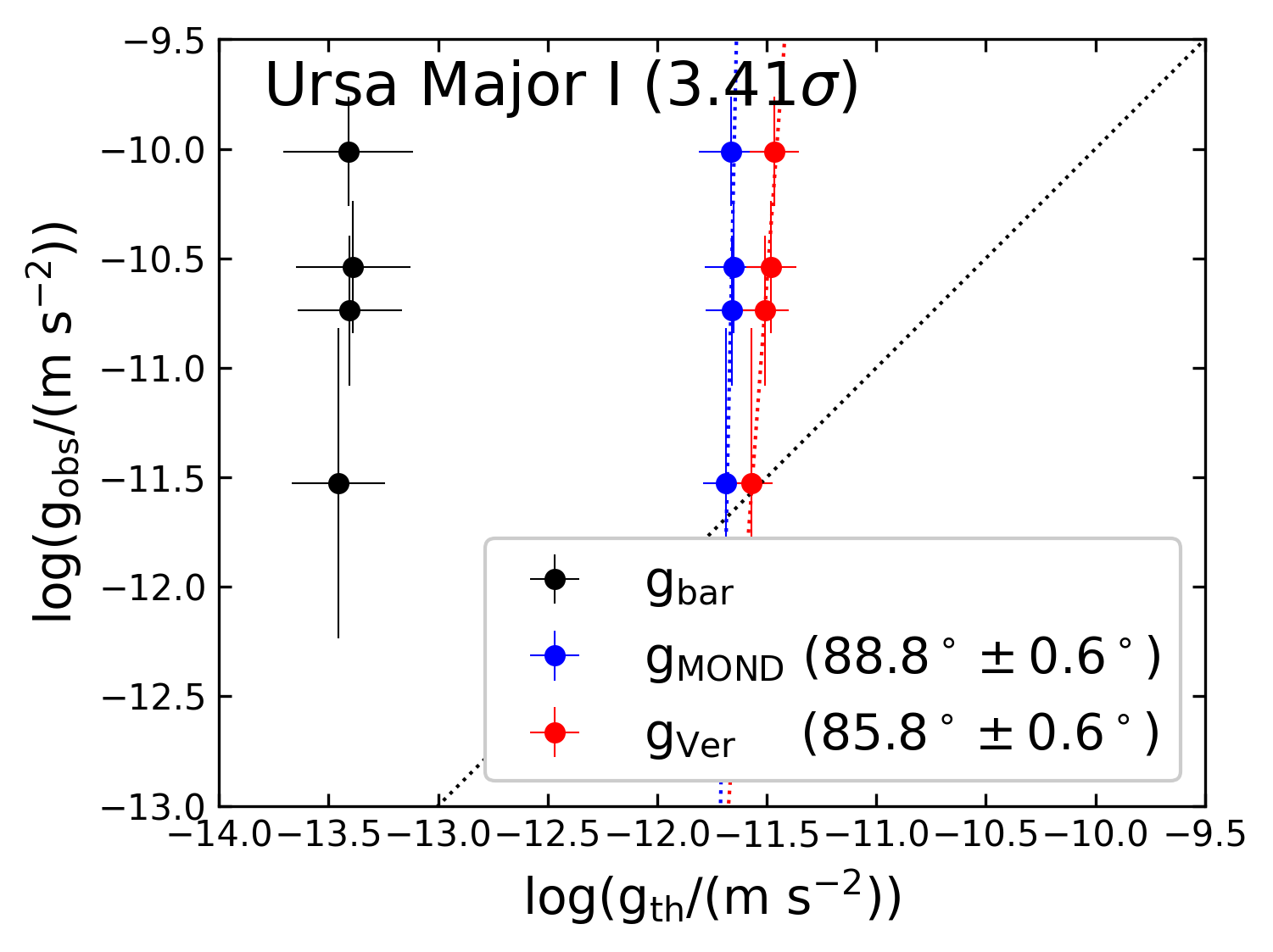} \\[1ex]

\hspace{0.4cm}\includegraphics[width=0.42\textwidth]{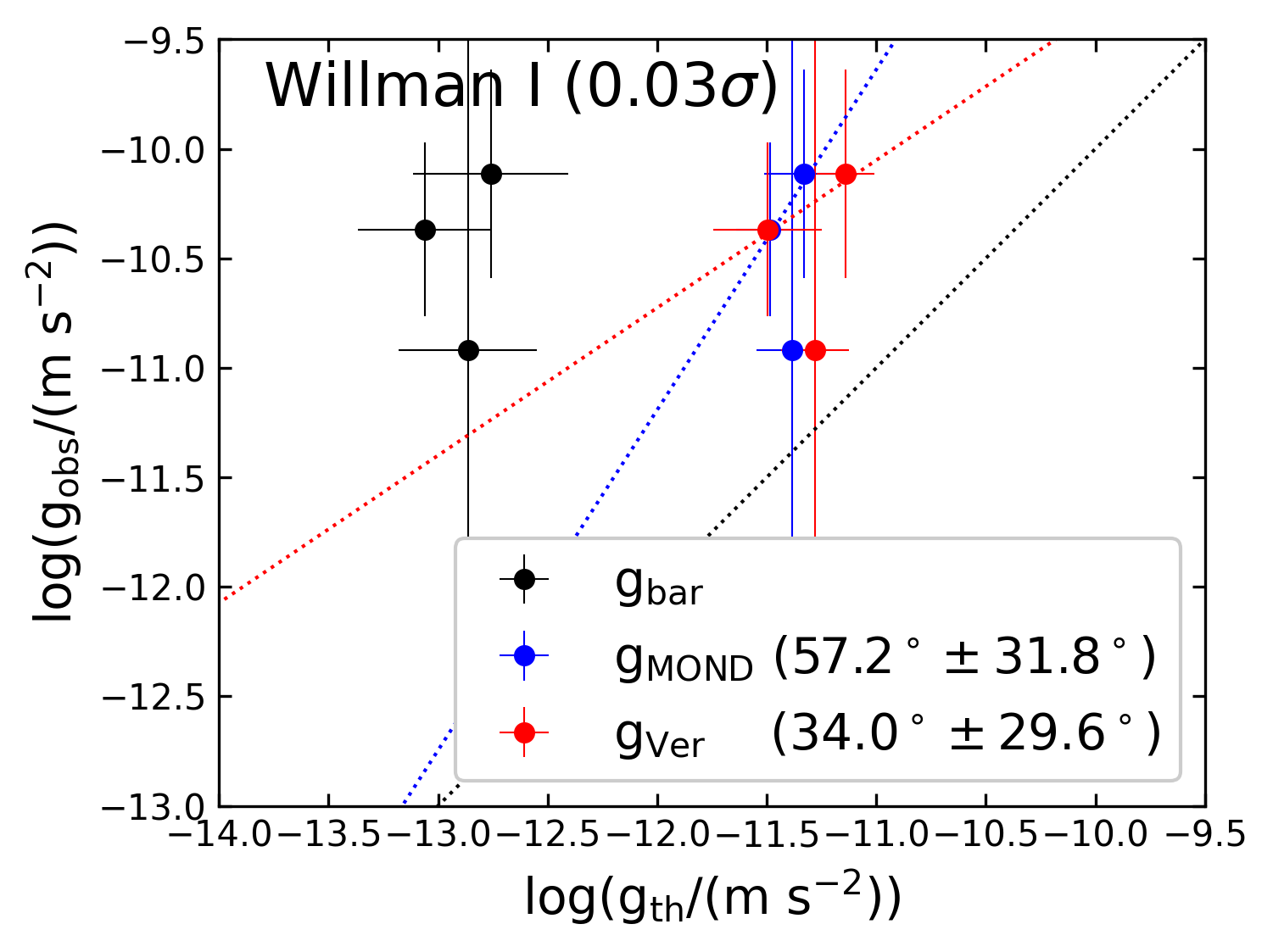} 

\caption{Same as Fig. \ref{eightfigures}, but for 7 other galaxies} \label{othersevenfigures}
\end{figure}

We analyze two groups of dwarf spheroidal galaxies. The first consists of 8 classical satellites of the Milky Way, as examined in \cite{Lelli:2016cui}; see Fig. \ref{eightfigures}. The second comprises 18 additional dwarf spheroidals; see Figs. \ref{othereightfigures} and \ref{othersevenfigures}. Although in our previous study we included the four dwarf spheroidals that have two data sets each—one to the left and one to the right of the center—we exclude them from the present analysis, primarily because they have substantial rotational velocity components that render the data unreliable. Furthermore, in the second group, we exclude the three dwarf spheroidal galaxies, Coma Berenices, Segue I, and Tucana, because each system provides only two data points, which renders the uncertainty on the fitted slope identically zero.

The summary of Figs. \ref{eightfigures}, \ref{othereightfigures}, and \ref{othersevenfigures} is listed in Table \ref{angle}. Among the remaining 23 samples, 21 favor Verlinde's emergent gravity, while 2 favor MOND. This indicates that Verlinde's emergent gravity reflects the observed data much better than MOND. Roughly half of the samples surpass a statistical significance of 1$\sigma$. When all results are combined using Stouffer's method, the overall significance reaches $5.2\sigma$, giving strong support to Verlinde’s emergent gravity. Fisher's method also yields a high statistical significance of $4.5\sigma$. More conservatively, disregarding the statistical significance of each sample, but considering only the fact that 21 out of 23 samples favored Verlinde's emergent gravity, we obtain $4.0\sigma$ from the binomial distribution.

\begin{figure}
\centering

\hspace{0.4cm}\includegraphics[width=0.42\textwidth]{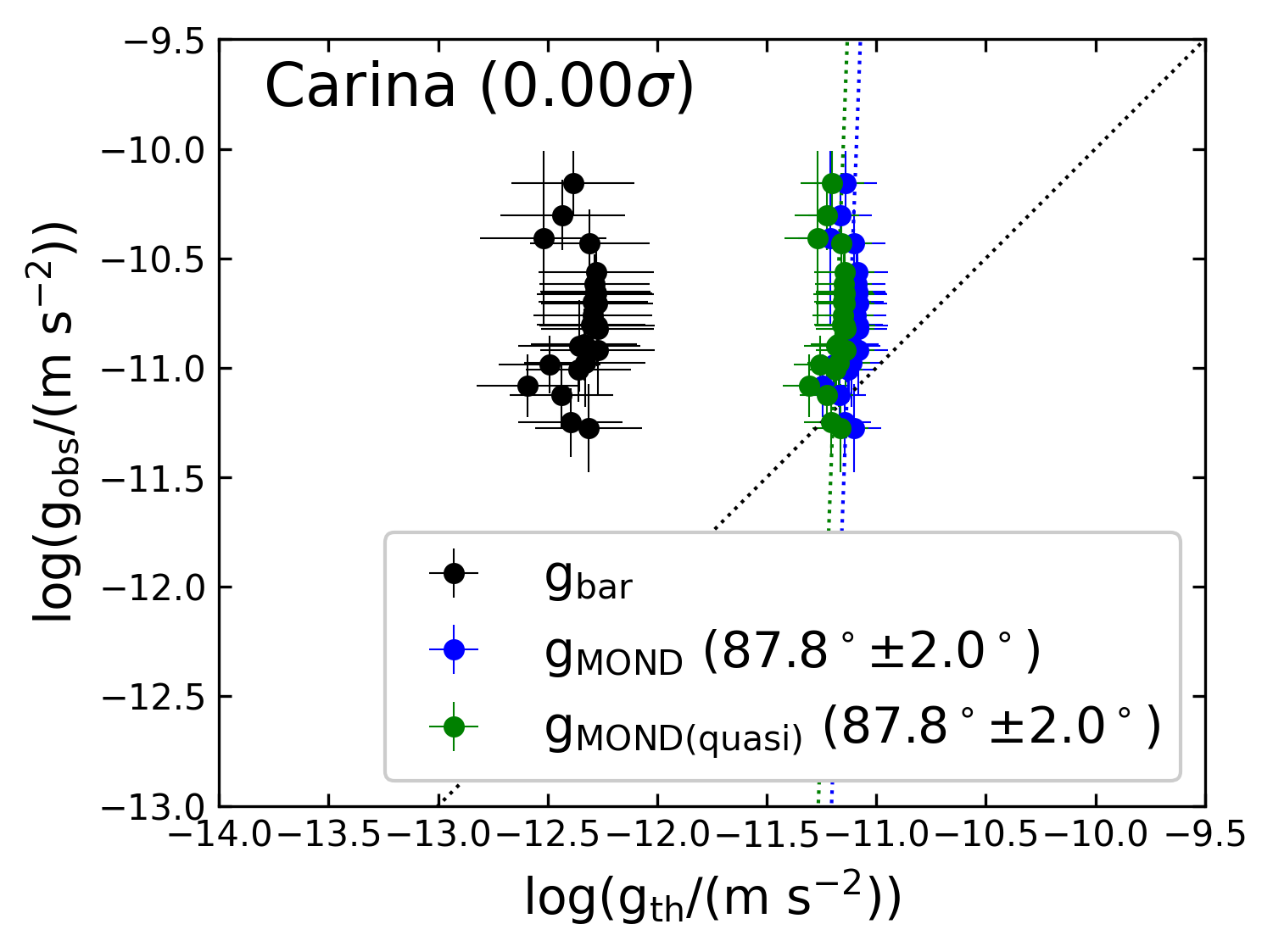} 
\includegraphics[width=0.42\textwidth]{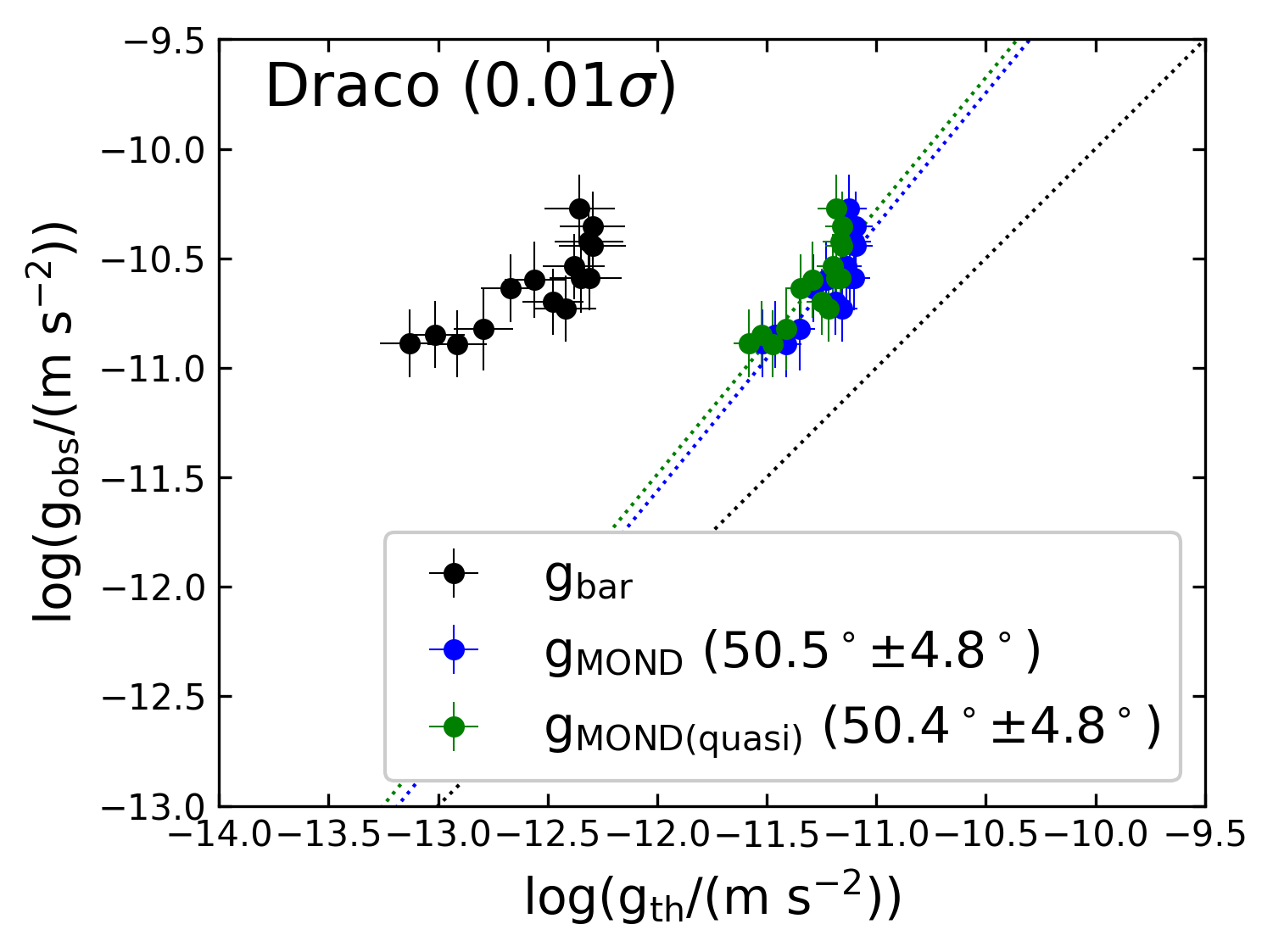} \\[1ex]

\hspace{0.4cm}\includegraphics[width=0.42\textwidth]{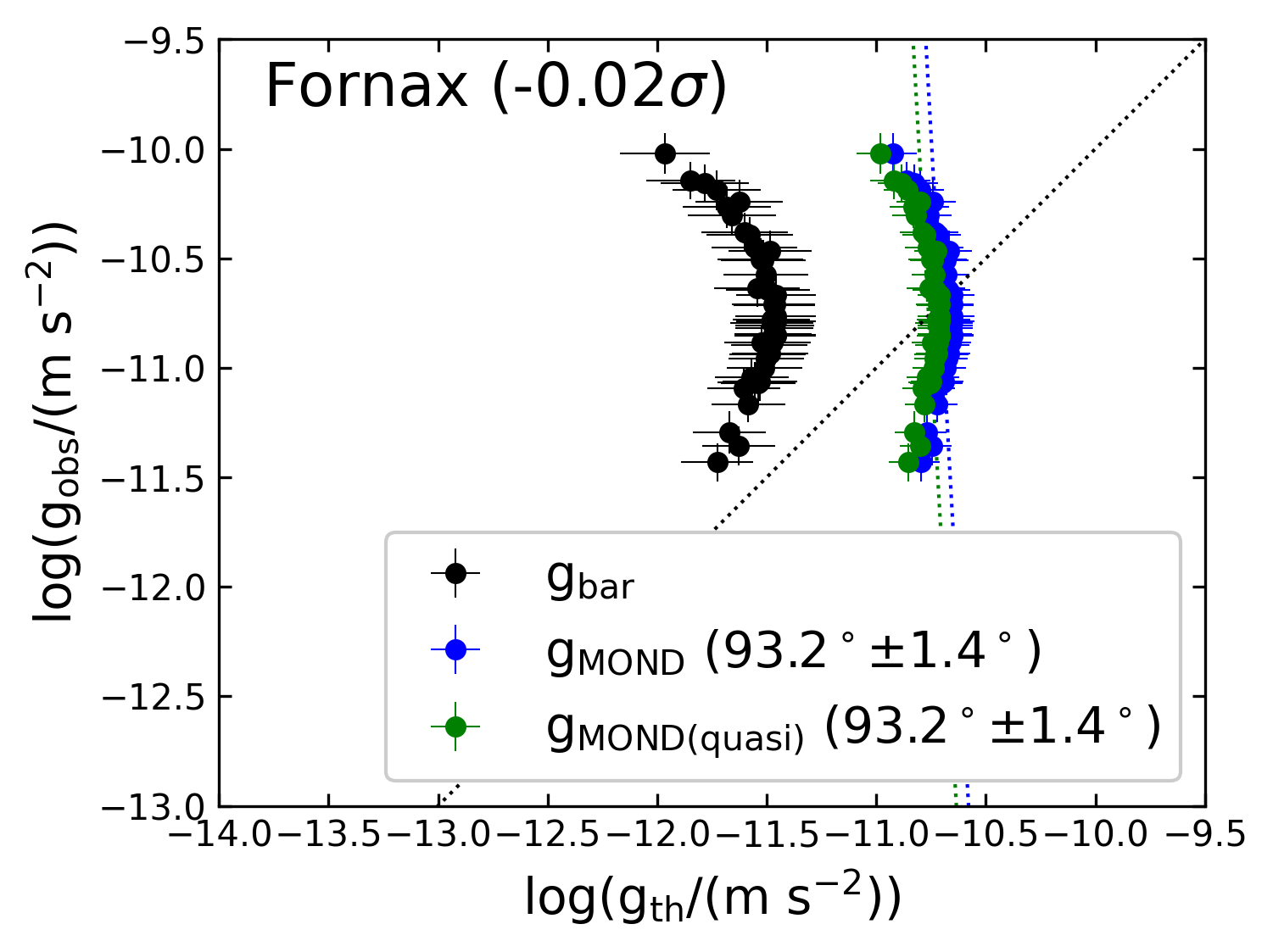} 
\includegraphics[width=0.42\textwidth]{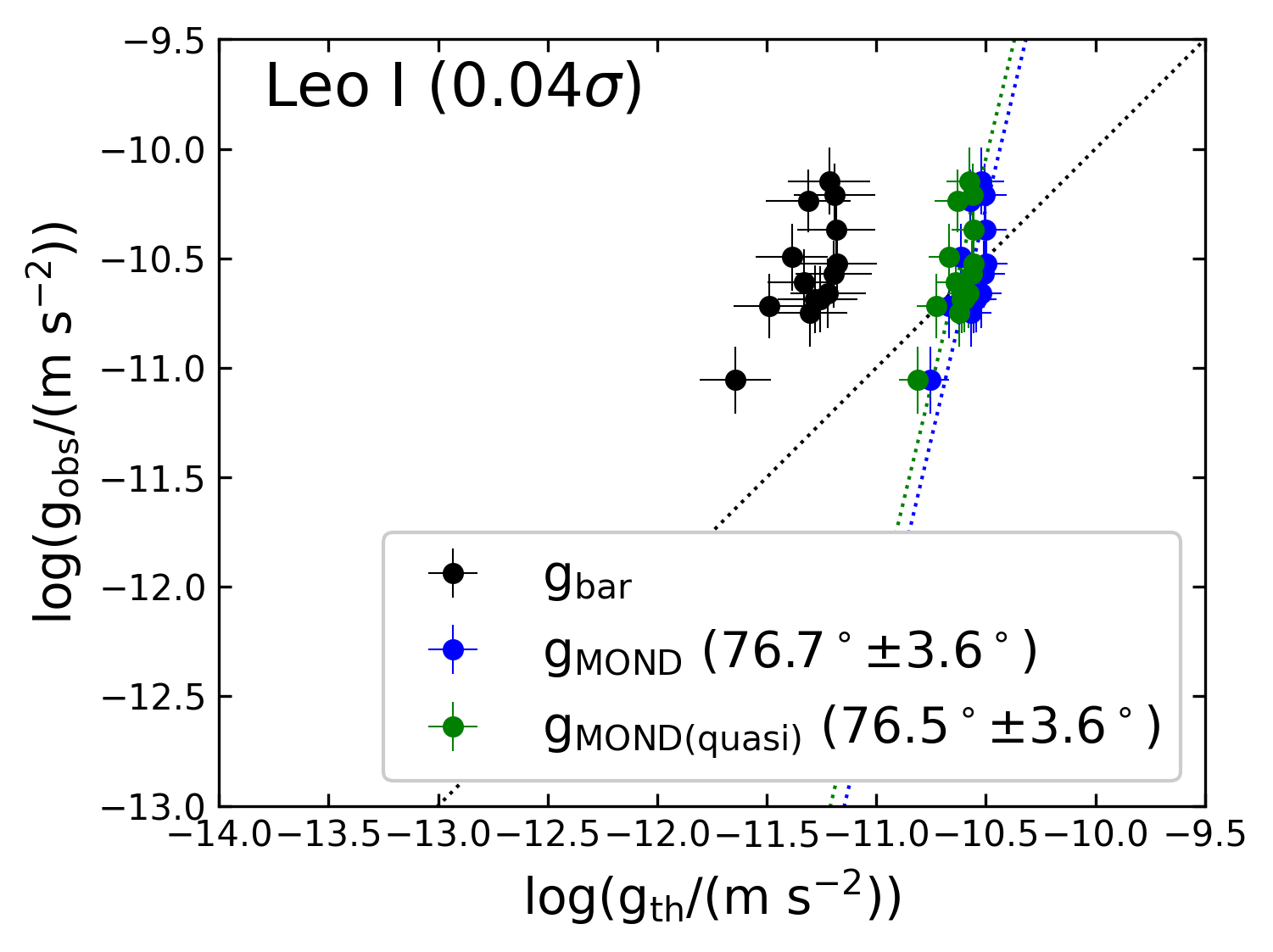} \\[1ex]

\hspace{0.4cm}\includegraphics[width=0.42\textwidth]{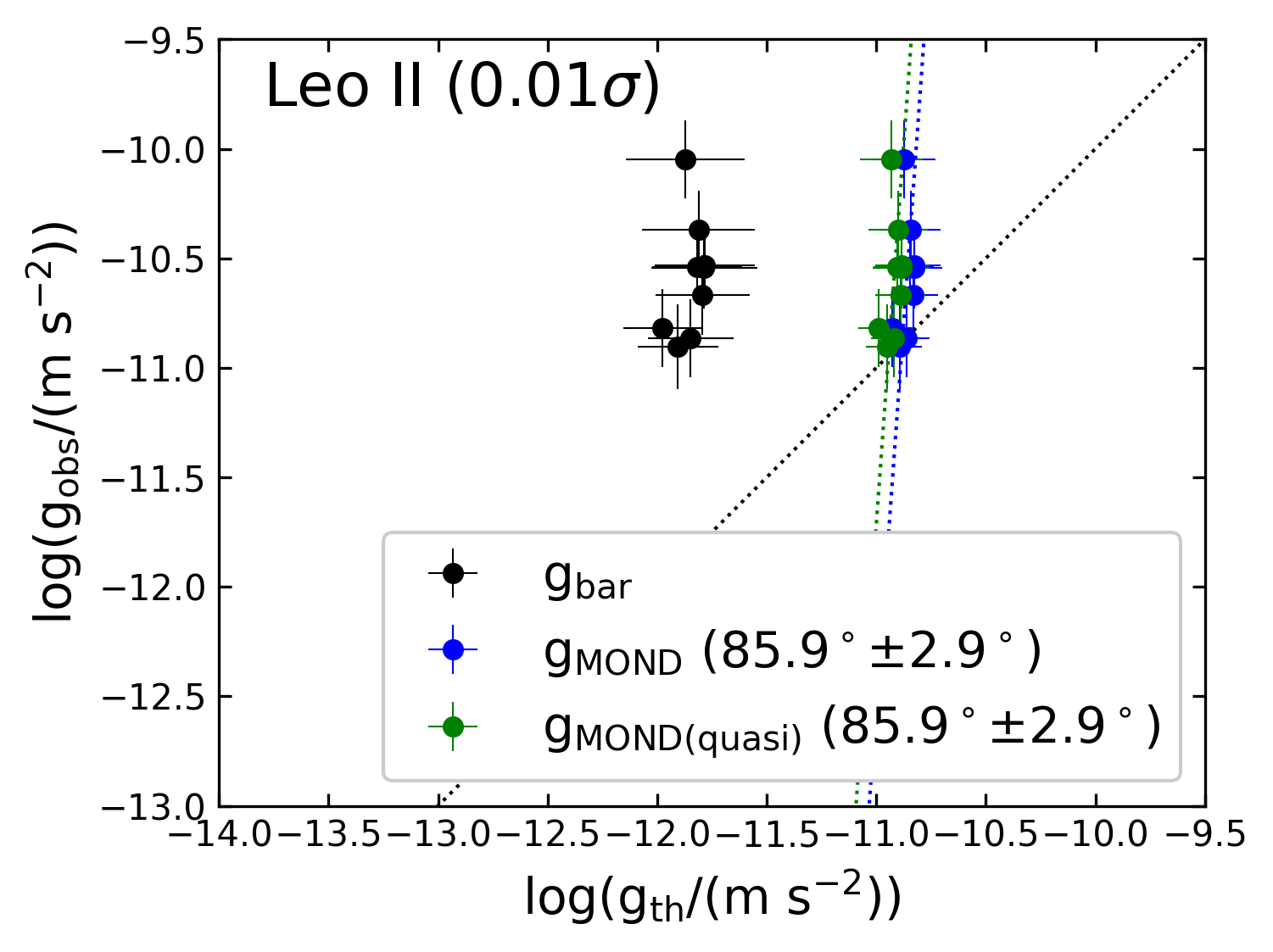} 
\includegraphics[width=0.42\textwidth]{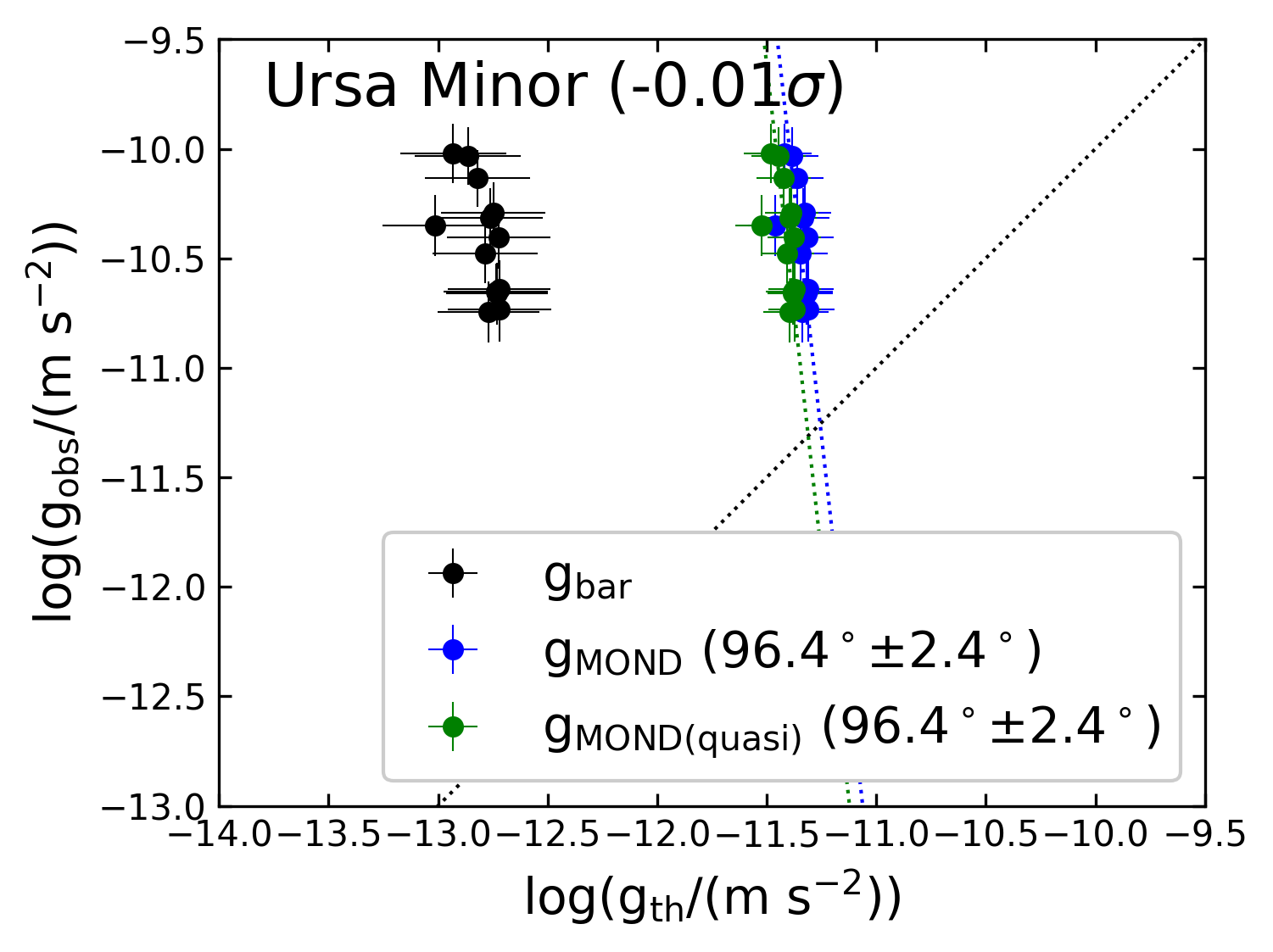} \\[1ex]

\hspace{0.4cm}\includegraphics[width=0.42\textwidth]{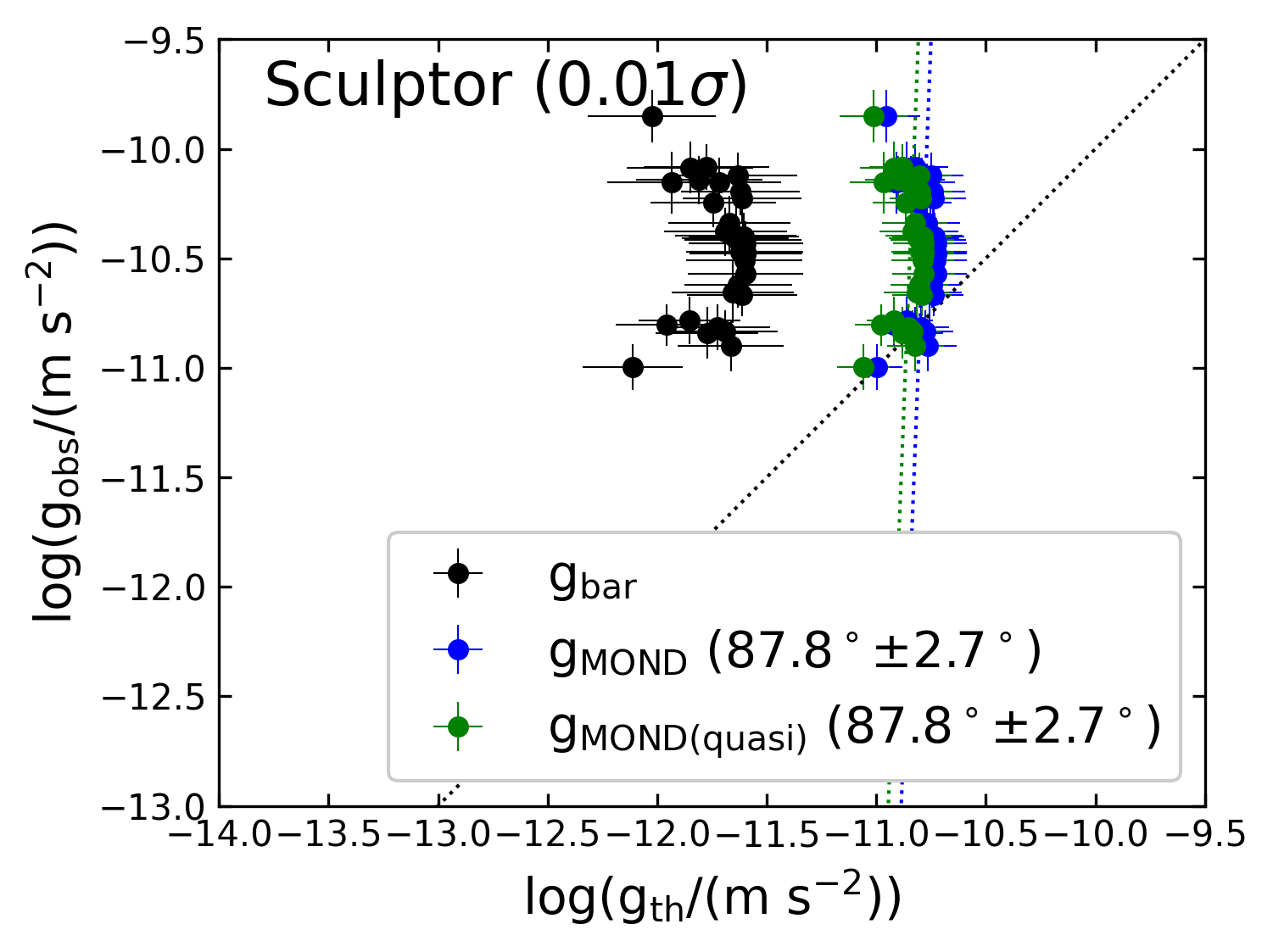} 
\includegraphics[width=0.42\textwidth]{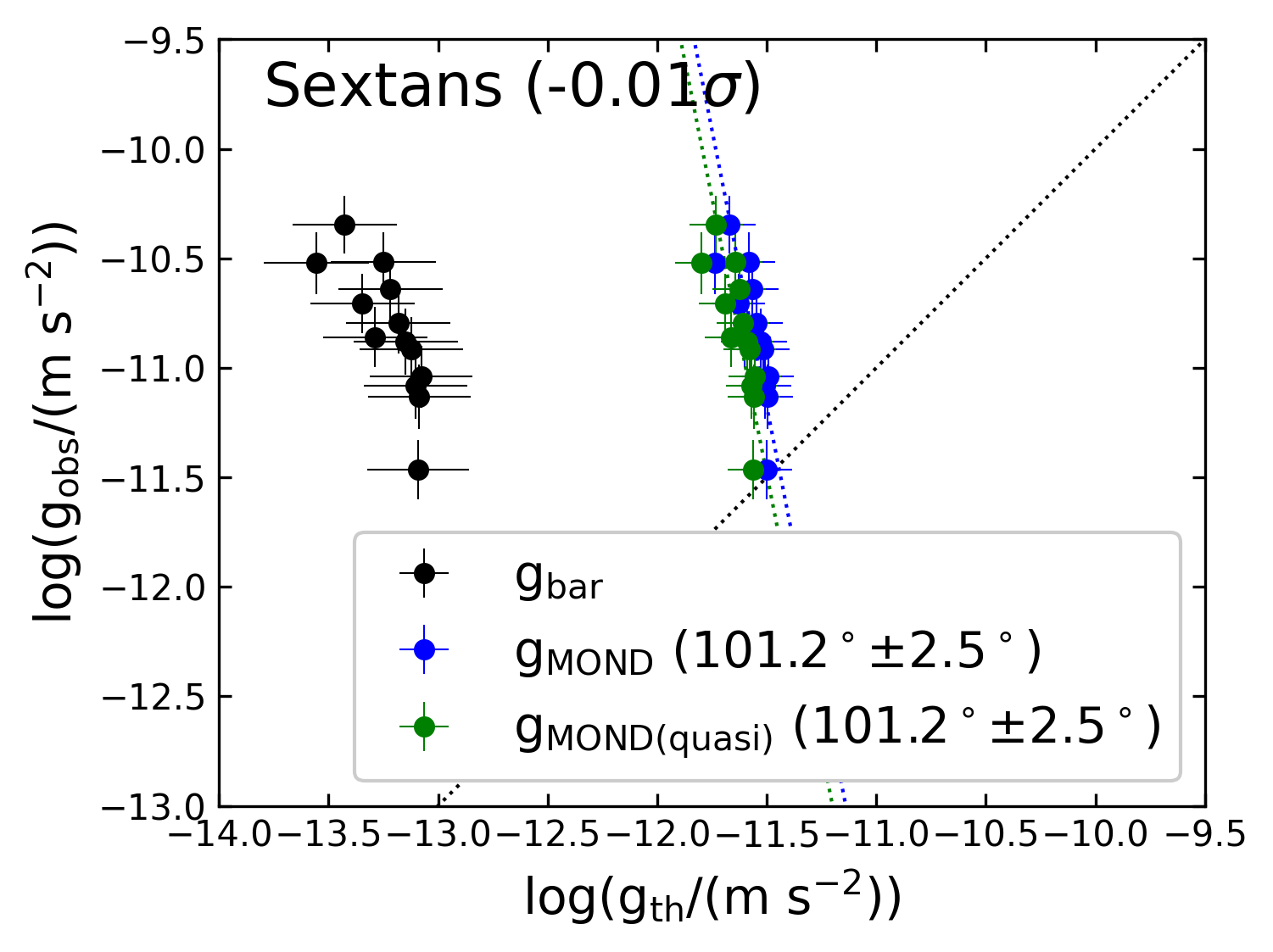}

\caption{Same as Fig. \ref{eightfigures}, except that the Verlinde prediction is replaced by the MOND prediction calculated with Milgrom's constant corresponding to the quasi-de-Sitter value of $a_0$. Changing Milgrom's constant does not significantly alter the trend, indicating that the behavior preferred by Verlinde’s emergent gravity is not simply a consequence of adopting a smaller value of $a_0$. In addition, $g_{\rm Ver}$ is generally larger than $g_{\rm MOND}$, whereas $g_{\rm MOND(quasi)}$ is smaller than $g_{\rm MOND}$. Thus, the trend characteristic of Verlinde's emergent gravity clearly arises from the specific functional form of its formula.} \label{quasifigures}
\end{figure}

At this stage, one might suspect that the preference for Verlinde's emergent gravity stems from adopting the quasi-de Sitter value of $a_0$ instead of the specific form of $g_{\rm Ver}$. For that choice, $a_0/6$ is approximately 25\% smaller than $a_M$. To examine this possibility, we recomputed the MOND prediction Eq.~(\ref{RAR}) for the value of Milgrom's constant $a_M = a_0/6$, with $a_0$ taken as the quasi-de Sitter value, and compared the outcome; see Fig.~\ref{quasifigures}. It is clear that lowering $a_M$ by 25\% barely affects the overall behavior. Furthermore, the earlier three figures show that $g_{\rm Ver}$ typically exceeds $g_{\rm MOND}$ even though $a_0/6$ is 25\% smaller than $a_M$. Although $g_{\rm MOND}$ would be reduced by using the smaller value of $a_M$, the ``Verlinde effect'' more than compensates for this reduction and shifts the trend in favor of Verlinde’s emergent gravity. In fact, under spherical symmetry, which we assume to exist in our dwarf spheroidal samples, Eq.~(\ref{gD2}) can be rewritten as
\begin{equation}
    g_d^2 = \frac{a_0}{6}\bigl(g_{\rm bar} + 4\pi G \rho_{\rm bar} r\bigr).\label{4piGrhobarr}
\end{equation}
Figs.~\ref{eightfigures}, \ref{othereightfigures} and \ref{othersevenfigures} make it evident how crucial the $4\pi G \rho_{\rm bar} r$ term is. It is not the smallness of the quasi-de Sitter value that makes Verlinde values follow the observed values more closely, but the specific form of Verlinde gravitational acceleration, especially the $4\pi G \rho_{\rm bar}$, which cannot be expressed directly as a function of $g_{\rm bar}$.

\begin{table}[h]
\captionsetup{width=14cm}
\centering

\begin{tabular}{|l|c|c|c|}
\hline
 & MOND($^\circ$) & Verlinde($^\circ$) & Stat. signific.\\
\hline
Carina       & 87.8$\pm$2.0 & 83.1$\pm$2.7 & 1.44$\sigma$\\
Draco        & 50.5$\pm$4.8 & 39.0$\pm$4.9& $-0.07\sigma$\\
Fornax       & 93.2$\pm$1.4& 88.5$\pm$1.7 & 2.14$\sigma$\\
Leo I        & 76.7$\pm$3.6& 66.9$\pm$5.0 & 1.59$\sigma$\\
Leo II       & 85.9$\pm$2.9& 77.8$\pm$4.2 & 1.60$\sigma$\\
Ursa Minor   & 96.4$\pm$2.4& 92.0$\pm$2.3 & 1.31$\sigma$\\
Sculptor     & 87.8$\pm$2.7& 85.6$\pm$2.9 & 0.55$\sigma$\\
Sextans      & 101.2$\pm$2.5& 98.9$\pm$2.7 & 0.63$\sigma$\\
And I & 92.3$\pm$1.3 & 93.1$\pm$2.6 &$-0.25\sigma$\\
And III & 94.8$\pm$1.4& 93.8$\pm$3.9
&0.24$\sigma$\\
And V & 88.5$\pm$3.0& 80.1$\pm$5.3
&1.37$\sigma$\\
And VI & 85.0$\pm$4.5& 73.9$\pm$10.4&0.99$\sigma$\\
And VII & 90.7$\pm$1.5& 86.5$\pm$2.6&1.38$\sigma$\\
And IX & 89.4$\pm$2.7& 82.3$\pm$5.3&1.20$\sigma$\\
And XIV & 93.8$\pm$2.3& 88.5$\pm$1.7&1.83$\sigma$\\
And XXI & 89.6$\pm$0.3& 88.1$\pm$1.4&0.99$\sigma$\\
And XXIII& 97.8$\pm$2.3& 92.9$\pm$4.6&0.94$\sigma$\\
Bootes I& 91.6$\pm$0.9& 91.3$\pm$0.7&0.25$\sigma$\\
Canes Venatici I& 102.8$\pm$2.8& 100.9$\pm$2.5&0.49$\sigma$\\
Cetus& 87.3$\pm$2.2& 79.5$\pm$3.7&1.81$\sigma$\\
Hercules& 90.5$\pm$2.0& 88.7$\pm$0.6&0.86$\sigma$\\
Ursa Major I& 88.8$\pm$0.6& 85.8$\pm$0.6&3.41$\sigma$\\
Willman I& 57.2$\pm$31.8& 34.0$\pm$29.6&0.03$\sigma$\\

\hline
\end{tabular}
\caption{Angles of the slopes of the lines fitting data. For all the spheroidals except two, the Verlinde case has a slope closer to 45$^\circ$, which implies that it fits the data better than MOND. The statistical significance of the Verlinde slope being closer to 45$^\circ$ is also provided. When all are combined, the overall statistical significance is 5.2$\sigma$.}
\label{angle}
\end{table}

\section{Conclusions}\label{conclusions}
In our earlier work \cite{HanHwangYoon}, we had applied Verlinde's emergent gravity and MOND to calculate gravitational accelerations in dwarf spheroidals, and compared them with the observed values to conclude that Verlinde values are closer to the observed values than MOND values are. In the present paper, instead of this simple comparison, we performed a more interactive comparison, examining how the theoretical and the observed accelerations evolve \emph{within} each spheroidal galaxy. Out of 23 spheroidal galaxies, we found that for 21 of them the Verlinde predictions follow the observed values more closely than the MOND prediction, the overall combined statistical significance being 5.2$\sigma$. We further verified that this behavior cannot be attributed to our choice of the quasi-de Sitter value for $a_0$, whose value $a_0/6$ is smaller than $a_M$. The fact that the Verlinde predictions are larger than the MOND predictions, even with this smaller quasi-de Sitter value, indicates that the particular form of the Verlinde gravity formula compensates in such a way that it yields values closer to, and more tightly following, the observations than the MOND predictions. The term in Eq.~(\ref{4piGrhobarr}), $4\pi G \rho_{\rm bar}r$, which cannot be expressed directly by $g_{\rm bar}$, plays a crucial role. 
 

To conclude, previous studies confirmed that Verlinde's emergent gravity and MOND both successfully account for the accelerations in rotation-supported galaxies remarkably well, a feat often attempted but never fully achieved in dark matter theory. Therefore, both theories seem to be on the right track, and our test of dwarf spheroidals has identified which of the two is superior.

\section*{Acknowledgement}

We acknowledge the support of the National Research Foundation of Korea (NRF) grant funded by the Korean government (MSIT), [NRF-2022R1A2C1092306(YY), NRF-2021R1A2C1094577(SH, HSH)] and Hyunsong Educational \& Cultural Foundation.

\end{document}